\documentclass{aa}
\usepackage{graphicx}

\newcommand{\mgn}{$^{\rm m}$}
\newcommand{\nh}{$N_{\rm H}$}

\newcommand{\col}{cm$^{-2}$}
\newcommand{\flux}{erg\,s$^{-1}$\,cm$^{-2}$}

\begin{document}
   \title{The Hamburg/RASS Catalogue of optical identifications}
   \subtitle{Northern high-galactic latitude ROSAT
   Bright Source Catalogue X-ray sources
   }

   \author{F.-J. Zickgraf
          \inst{1}  \and D. Engels  \inst{1} \and  H.-J. Hagen \inst{1}
 \and  D. Reimers \inst{1} \and  W. Voges  \inst{2}
          }
   \institute{Hamburger Sternwarte, Gojenbergsweg 112, 21029 Hamburg, Germany
       \and Max--Planck--Institut f\"ur extraterrestrische Physik,
        Postfach 1312, 85741 Garching, Germany
}
   \offprints{F.-J. Zickgraf}
   \date{Received 02-08-02; accepted 06-05-03}

\abstract{
We present the Hamburg/RASS Catalogue (HRC) of optical identifications of 
X-ray sources at high-galactic latitude. The HRC includes all X-ray sources 
from the ROSAT Bright Source Catalogue (RASS-BSC) 
with galactic latitude $|b| \ge 30\degr$ and declination
$\delta \ge 0\degr$. 
In this part of the sky covering $\sim$10\,000\,deg$^2$ the RASS-BSC 
contains  5341 X-ray sources. For the optical identification we used  
blue Schmidt prism and direct plates taken for the  northern
hemisphere Hamburg Quasar Survey (HQS) which are now available in digitized form.
The limiting magnitudes are 18.5 and 20, respectively.
For 82\%  of the selected RASS-BSC an identification could be given. For the rest
either no counterpart was visible in the error circle or a plausible
identification was not possible. With $\sim$42\% AGN represent the largest 
group of X-ray emitters, $\sim$31\% have a stellar counterpart, whereas
galaxies and cluster of galaxies comprise only $\sim$4\% and $\sim$5\%, 
respectively. In $\sim$3\% of the RASS-BSC sources no object was visible on 
our blue direct plates within 40\arcsec\ around the X-ray source
position. The catalogue is used as a source for the selection of (nearly) 
complete samples  of the various classes of X-ray emitters.
   }
\titlerunning{The Hamburg/RASS  Catalogue of optical identifications}
\authorrunning{F.-J. Zickgraf et al.}
\maketitle

\section{Introduction}
The ROSAT All-Sky Survey (RASS) was performed in 1990/91 
within the first six months of the ROSAT X-ray satellite mission 
(Tr\"umper \cite{Trumper83}). It was the first all-sky survey in
the soft X-ray range from 0.07 keV to 2.4\,keV with an imaging telescope
(Aschenbach \cite{Aschenbach88}).
The results of the survey, 
photon data as well as source catalogues, have been made available to the community. 
In total nearly 125\,000 X-ray sources were detected and 
published in two  catalogues. The ROSAT Bright Source Catalogue (RASS-BSC), 
contains 18\,811 ROSAT all-sky survey sources with  count-rates  $\ge 0.05$\,cts\,s$^{-1}$ in the 0.1-2.4\,keV energy range
(Voges et al. \cite{Vogesetal99}). The ROSAT 
Faint Source Catalogue (RASS-FSC) comprises  105\,924 X-ray sources 
with count rates below 0.05\,cts\,s$^{-1}$ (Voges et al. \cite{Vogesetal00}).

For follow-up studies the nature of the sources must be known in
order to select objects of the class to be studied from the RASS source list.
The ROSAT data alone usually do not allow 
one 
to identify the nature of an X-ray source.
Additional information, in particular from optical observations, is necessary. 
In the past decade numerous programs for the optical
identification of RASS X-ray sources have been performed. Many of these programs aimed at
the identification of objects belonging to certain classes of X-ray emitters by applying 
preselection criteria, like X-ray count rates, hardness ratios and source extension. 
Other programs aimed at a statistical identification of the RASS without preselection 
on the basis of X-ray properties. 
Two studies of this type reaching
the RASS detection limit are the
ROSAT North Ecliptic Pole Survey (Henry et al. \cite{Henryetal01}, Voges et al. 
\cite{Vogesetal01}) and the 
LSW-INAOE-MPE project (Zickgraf et al. \cite{Zickgrafetal97}, Appenzeller et al.
\cite{Appenzelleretal98}, Krautter et al. \cite{Krautteretal99}).  
In these studies the sky area and the number of
sources were relatively small, $\simeq$ 1-2\% of the sky and  
$\la 1$\% of the RASS sources, respectively. 
An investigation covering a large area of sky is the  ROSAT Bright Survey 
(RBS) of $\sim$2000 X-ray bright 
RASS sources with count rates of $>0.2$\,cts\,s$^{-1}$ (Fischer et al. 
\cite{Fischeretal98}, Schwope et al. \cite{Schwopeetal00}).

The nature of the vast majority of RASS sources is still unknown. 
The  potential of the RASS  lying in  the all-sky coverage 
with high sensitivity
has  still not been fully exploited. 
This is  particularly important in searching 
for rare objects which require one to survey large areas of the sky. 
The optical identification process 
can be made significantly more efficient if a preselection of the possible optical 
counterparts is available. The Hamburg/RASS optical identification 
project aims at the identification of all RASS sources in the northern extragalactic 
sky as a basis for detailed follow-up studies. This goal can be reached by using 
Schmidt prism plates covering large areas of sky. In this way we
have finished the identification of all RASS-BSC sources with 
$\delta \ge 0\degr$ and galactic latitudes $\mid b\mid \ge 30\degr$.  In this region 
of the sky 
covering 10\,313\,deg$^2$
the BSC contains 5341 X-ray sources.

A first version of the Hamburg/RASS Catalogue of optical identifications
(HRC) containing identifications of 4190 RASS-BSC sources in the  sky region 
defined above plus additional sources in the galactic latitude range 
$20\degr \le |b| \le 30\degr$
was published by Bade et al. (\cite{Badeetal98a}) (Paper I hereafter). 
Here we present the full catalogue of the complete sample of 5341 RASS-BSC sources. 
The catalogue itself is available 
electronically\footnote{http://www.hs.uni-hamburg.de/hrc.html}. 
In Sect. \ref{technique} the identification 
technique is summarized. Identification criteria are discussed in Sect. 
\ref{criteria}. The catalogue is described in Sect. \ref{description}. 
A statistical analysis is given in Sect. \ref{discussion} and finally the results 
are summarized in Sect. \ref{sum}.

\section{Identification technique}
\label{technique}

The identification of the RASS-BSC sources was performed in two steps. First, the
optical sources in the error circles of the RASS sources were classified. Then the 
optical and the X-ray data were combined to assign the most likely counterpart to the 
X-ray source. 

The optical classification is based on 567 blue Schmidt prism plates  
and a set of blue direct plates collected during the  Hamburg Quasar Survey 
(HQS) covering the 
northern high galactic latitude sky with $|b| > 20\degr$  
(Hagen et al. \cite{Hagenetal95}). 
Because of 
the 
strong increase in the overlap rate of spectra towards low galactic
latitude we restricted the RASS identification to galactic latitude 
$|b| > 30\degr$. 
The average limiting magnitudes are $B_{\rm lim, p} = 18.5$ for the
prism plates and $B_{\rm lim, d} = 20$ for the direct plates.
The identification technique used for the first version of the HRC is
described in detail in Paper I. 
Although since then some modifications of the procedure have been implemented, 
the general identification process for the remaining $\sim1200$ sources 
was the same as for the first part of the HRC with the exception of 
the scanning procedure.
In particular, we have now at our disposal digitized high resolution
scans of one of the two Schmidt prism plates obtained for each field. 
The identification procedure for the new sources was as follows.
We first searched on the digitized direct plates for all possible counterparts 
in the X-ray error circle. A search radius of 60\arcsec\ was adopted. Note that this
is significantly larger then 
the nominal ROSAT 3$\sigma$ error circle which is  $\approx$40\arcsec\ 
(Voges et al. \cite{Vogesetal99}). 
For a few Schmidt prism fields no direct plates were available. 
In these cases we used the Digitized Sky Survey instead. Then spectra were extracted 
from the data base of the digitized spectral plates at the positions of the objects 
instead of scanning the plates individually as described in Paper I.
Using these spectra the objects were classified interactively. Object 
classes used for the classification are listed in Table \ref{category}.
In Sect. \ref{criteria} more details on classification criteria are given. 
Finally, all objects were entered into the catalogue and 
the most likely  counterparts to the X-ray sources were identified by a flag. 
For the source identification the codes and object categories listed in Table \ref{idcode} 
were used. 
Object classes, ID codes and object categories are the same as in Paper I. 
We then created the final catalogue by extracting the 4190 entries from the 
first version of the  HRC and adding the missing 1151 source identifications.

\begin{table*}
\caption[]{Object classes used for the classification of the prism spectra 
of objects found in the X-ray error circles.}   
\begin{tabular}{ll}
\noalign{\smallskip}
\hline
\noalign{\smallskip} 
class & description\\
\noalign{\smallskip}
\hline
\noalign{\smallskip} 
W-DWARF & white dwarf, blue continuum, broad absorption lines\\
CV & cataclysmic variable, exhibits Balmer emission\\
STAR-BA & Balmer line absorption\\
STAR-FG & G-band, Ca\,H\&K absorption\\
STAR-K & G-band, Ca\,H\&K absorption, redder continuum than STAR-FG\\
STAR-M & continuum very red, TiO and CaI\,4226\AA\ absorption\\
GALAXY & red continuum, no emissison lines, extended direct image\\
BLUE GALAXY & continuum blue, extended direct image\\
AGN & continuum blue, emission lines\\
QSO & continuum very blue, point-like image, emission lines \\
EBL-WK& continuum very blue, faint point-like image\\
BLUE-WK & blue continuum, faint point-like image\\
RED-WK & red continuum, faint point-like image\\
UNIDENT & none of the above defined classes assignable\\
OVERLAP &  no classification possible due to overlapping spectra\\
SATURATE &  no classification possible due to saturated spectrum\\
\noalign{\smallskip}
\hline
\noalign{\smallskip}
\end{tabular}
\label{category}
\end{table*}

\section{Classification criteria and object categories}
\label{criteria}
A detailed description of the classification criteria has been given in Paper~I. 
Here we shortly summarize the classification criteria of the various spectral classes.
Sample spectra are shown in the Appendix.

Spectral classification was performed on the basis of the appearance of 
both the
high-resolution density spectra and the images on the direct plates.  
The majority of optical objects in the RASS error circles were point sources. 
Objects with blue continua and occasionally emission lines dominated. Most of them
are Seyfert galaxies or QSO. 
A minority of blue objects was classified as cataclysmic variable. 
They show the typical Balmer and He emission lines whereas AGN
exhibit only few and usually  broad emission lines. 
A few blue objects showed the spectral features of white dwarfs. 
Other stellar sources were classified according to the spectral features listed in
Table \ref{category}. Spectra of stars brighter than $\sim12^{\rm m}-13^{\rm m}$ 
are saturated. Hence a spectral type could not be determined and they were  
classified as ``SATURATE''. 

The classification as 
a galaxy 
required an extended direct image of the optical source and the spectral features of
galaxies, in particular the presence of the Ca break. A subclass of galaxies showed a 
conspicuous blue continuum, leading to the classification as ``BLUE GALAXY''.

Optically faint objects could not be classified into specific classes  due to the low
S/N ratio of the spectra. For these spectra three categories were used corresponding
to the continuum properties. The class ``EBL-WK'' is characterized by weak but extremely 
blue spectra. The opposite are objects with weak red spectra. They were classified 
``RED-WK''. Intermediate cases were  classified as ``BLUE-WK''.  
Most EBL-WK and BLUE-WK  are
Sy 1 galaxies or QSOs. This was verified by slit spectra obtained during follow-up observations,
e.g. Engels et al. (\cite{Engelsetal98}).

Likewise, BL Lac objects could not individually be classified as such with 
our prism spectra. 
They are expected to appear mostly in the class of AGN, QSO, 
EBL-WK or BLUE-WK due to their usually blue continua. However, a small fraction probably 
occurs in the class of RED-WK as in few cases BL Lac spectra were found to be rather red 
(e.g. Perlman et al. \cite{Perlmanetal96}).

In the second step the spectral classifications were used to assign an 
object category to the counterpart of the X-ray source. These categories are 
designated by one of the identification codes listed in Table \ref{idcode}. 
For each identification a reliability flag was added which can take the values 0, 1 or 2 for 
``highly probable'', ``probable'', and  ``possible'' identifications, respectively (cf.
Paper~I). The identification process included the valuation of the distance 
between X-ray and optical source, X-ray
hardness ratios, X-ray source extension, and X-ray to optical flux ratios. 

X-ray sources with 
counterparts
of classes EBL-WK or BLUE-WK usually were combined 
with Sy galaxies and QSOs in a category designated ``AGN'' with ID code 1. 
It comprises Sy, QSOs, and BL Lacs. 
Galaxies and blue galaxies are summarized in category "galaxy" with ID code 2.
The identification of a galaxy 
cluster (ID code 3) was  based on the  presence  of 
several galaxies in the surroundings of the X-ray source position 
(cf. Paper I). X-ray properties like hard spectra or source extent often helped to 
improve the plausibility of the identification of an X-source with a galaxy cluster. 
For optically faint clusters there is some chance for misclassification as individual galaxy
if a bright cD galaxy is present but most other members are below the plate limit of the 
blue direct plates (see also Sect. \ref{sourcestat}).
Stars were identified according to their spectral class except the class
``SATURATE'' which were identified as ''bright star'' with ID code 7\underline{x}4. 
Sources with no (unambiguous) identification or only RED-WK objects in the 
error circle were considered  unidentified with ID code 8. 
X-ray sources with a unique optical counterpart on
the direct plates within $\approx$30\arcsec\  of the X-ray position, but fainter as the
spectral plate limit, were separated using an ID code 803.
In a considerable number of cases no object was visible on our blue 
plates within 40$\arcsec$ around the RASS positions. These sources were 
designated ``blank fields'' and identified with ID code 0.

\begin{table}
\caption[]{Codes and object categories used in the HRC for the identification of  X-ray sources. 
The identification reliability flag \underline{x} can take the values 0 = ``highly 
probable'',
 1 = ``probable'', 2 = ``possible''.}   
\begin{tabular}{ll}
\noalign{\smallskip}
\hline
\noalign{\smallskip}
ID code & object category\\
\noalign{\smallskip}
\hline
\noalign{\smallskip}
1\underline{x} & active galactic nucleus (``AGN'')\\
               & including Sy, QSO and BL Lac\\
2\underline{x} & galaxy\\
3\underline{x} & galaxy cluster\\
5\underline{x} & M star\\
6\underline{x} & white dwarf\\
7\underline{x}1 &K star\\
7\underline{x}2 &F-G star\\
7\underline{x}3 &cataclysmic variable\\
7\underline{x}4 &bright star\\
8 & no (unambiguous) identification, overlaps\\
803 & no spectrum available for likely counterpart\\
0 & no objects visible within 40''  around the X-ray \\
  & position (``blank field'')\\
\noalign{\smallskip}
\hline
\noalign{\smallskip}
\end{tabular}
\label{idcode}
\end{table} 

\section{Description of the catalogue}
\label{description}
All information was 
collected in a catalogue containing X-ray and optical data 
together with the source identification. In Fig. \ref{catalog} a section from the
catalogue is displayed as 
an 
example. Each entry consists of at least 
3 lines with X-ray data and  data of the optical sources in the error circle. 
The optical source identified 
as the counterpart 
of the RASS source is flagged by a 
``+'' sign 
preceding 
the running number of the optical source. In the case of blank fields the entry for
optical sources is absent and therefore only two lines are displayed. The contents 
of the individual lines and columns is as follows:

\begin{enumerate}
 \item The first line of each entry contains the X-ray data from the BSC 
 (cf. Voges et al. \cite{Vogesetal99}):
 \begin{enumerate}
  \item 1RXS designation
  \item count rate and error in cts\,s$^{-1}$
  \item hardness ratios with errors $HR1$, $\sigma (HR1)$, $HR2$, $\sigma (HR2)$
  \item source extension and extension likelihood
 \end{enumerate}
 \item second line:
 \begin{enumerate}
  \item RASS name as used in the first catalogue version
  \item right ascension and declination (J2000.0) of the X-ray source
  \item nominal 90\% error circle as provided by the SASS processing
  \item internal identifier of the HRC data base
  \item identification code of the source according to Table \ref{idcode}
  \item number of modifications of the identification
  \item date of identification/last modification
 \end{enumerate}
 \item third and following lines:
 \begin{enumerate}
  \item running number of objects visible in the error circle
  \item right ascension and declination (J2000.0) of the optical source
  \item positional offset  in arcsec between X-ray and optical source 
  in right ascension and declination
  \item distance between X-ray and optical source in arcsec
  \item $B$ magnitude
  \item $B-R$ colour from USNO-A\,2.0 catalogue (Monet et al.
  \cite{usno}) 
  \item spectral class according to Table \ref{category}
 \end{enumerate}
\end{enumerate}

$B$ photometry was usually taken from the HQS. 
The HQS $B$ magnitudes are calibrated in the Johnson system (Engels et al.
\cite{Engelsetal94}).
In cases of bright
stars, however, the  HQS photometry based on the low-resolution scans of the 
prism plates did not allow to measure the $B$ magnitude because of saturated 
spectra. We then searched the USNO
catalogue for the corresponding entry and adopted the USNO $B$ magnitude if a matching
entry  was found within 5 arcsec.
Due to proper motion effects this was not always the case. We then enlarged the search
radius to 10 arcsec. In this case matching objects were flagged by the letter  ``p'' following 
the $B$ magnitude. If no match was found even within 10 arcsec  only
a limit of ``$<10.0$'' is given for the respective object magnitude. 

Occasionally source confusion caused problems in the automatic assignment of 
$B$ magnitudes from the HQS for a faint object located close to brighter objects. 
In order to detect these cases we compared 
our $B$ magnitudes 
with that 
in the USNO catalogue and substituted USNO values if the HQS magnitude was  
brighter by at least 0.7 magnitudes. 
These magnitudes are flagged by the letter  ``u''. 
A correction of +0.4 was applied to the USNO  magnitudes in order to  transform the
photographic $O$ magnitude of the USNO to Johnson $B$ (Mickaelian et al.
\cite{Mickaelianetal01}).
The $B-R$ colours included in the HRC 
are 
all taken from the USNO catalogue. 

\begin{table*} 
\caption[]{Section of the catalogue. 
}
\begin{tabular}{l}
\noalign{\smallskip}
\hline
\noalign{\smallskip}
\end{tabular}
\centering
\includegraphics[width=17cm,angle=0,bbllx=85pt,bblly=200pt,bburx=565pt,bbury=540pt,clip=]{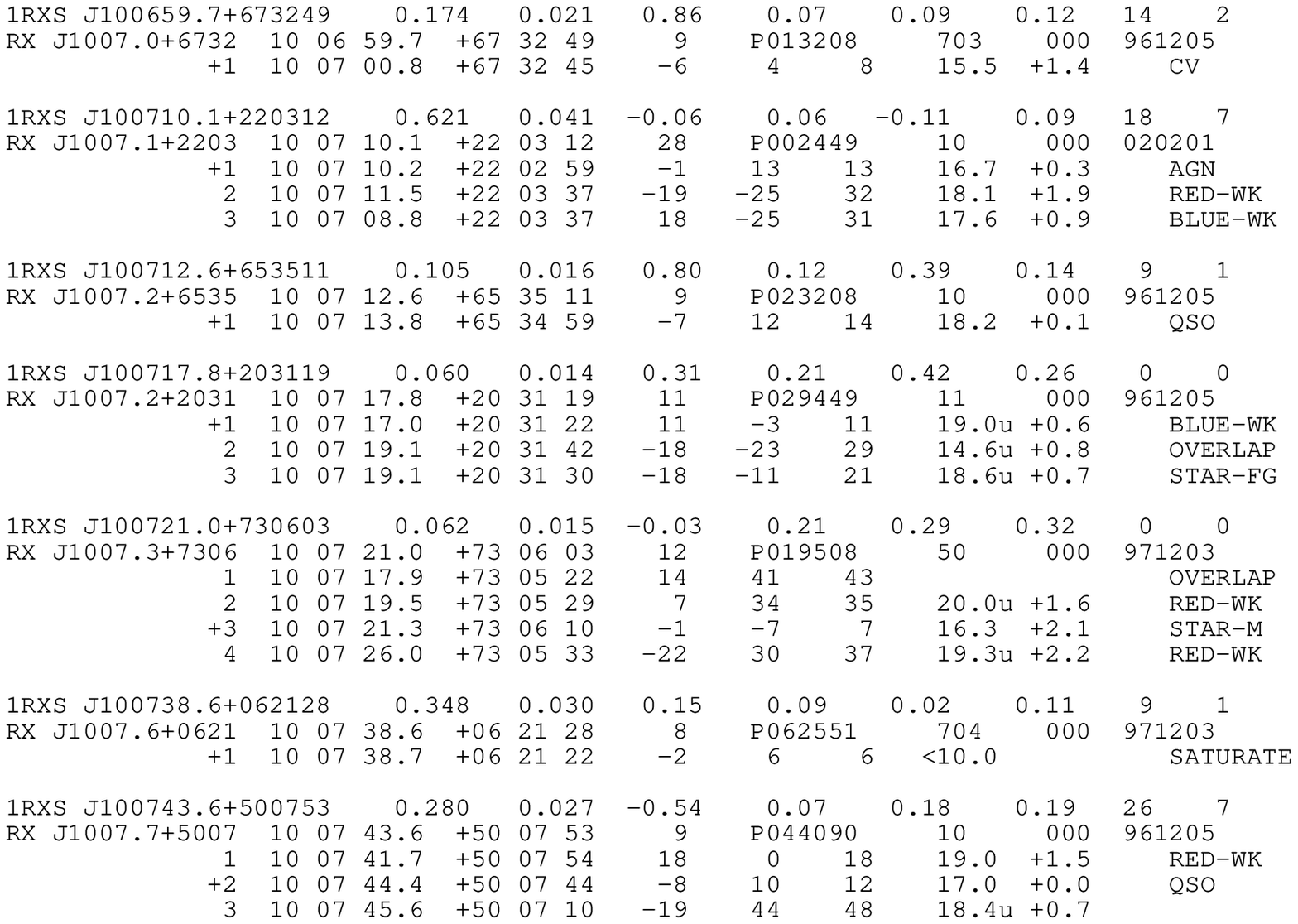}
\label{catalog}
\end{table*}

\begin{figure} 
\resizebox{\hsize}{!}{\includegraphics[angle=-90,bbllx=50pt,bblly=40pt,bburx=570pt,bbury=580pt,clip=]{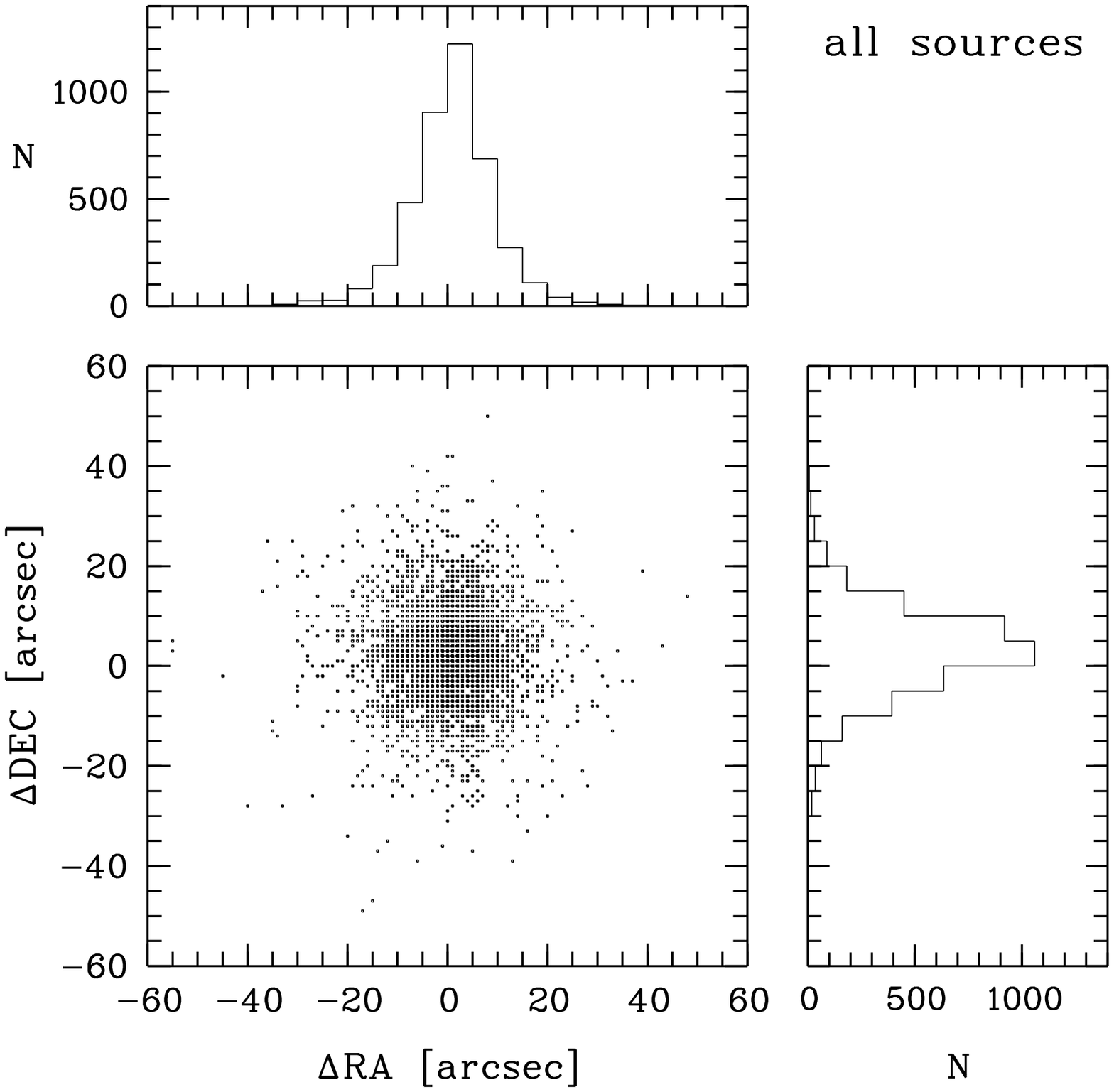}}
\resizebox{\hsize}{!}{\includegraphics[angle=-90,bbllx=150pt,bblly=60pt,bburx=510pt,bbury=545pt,clip=]{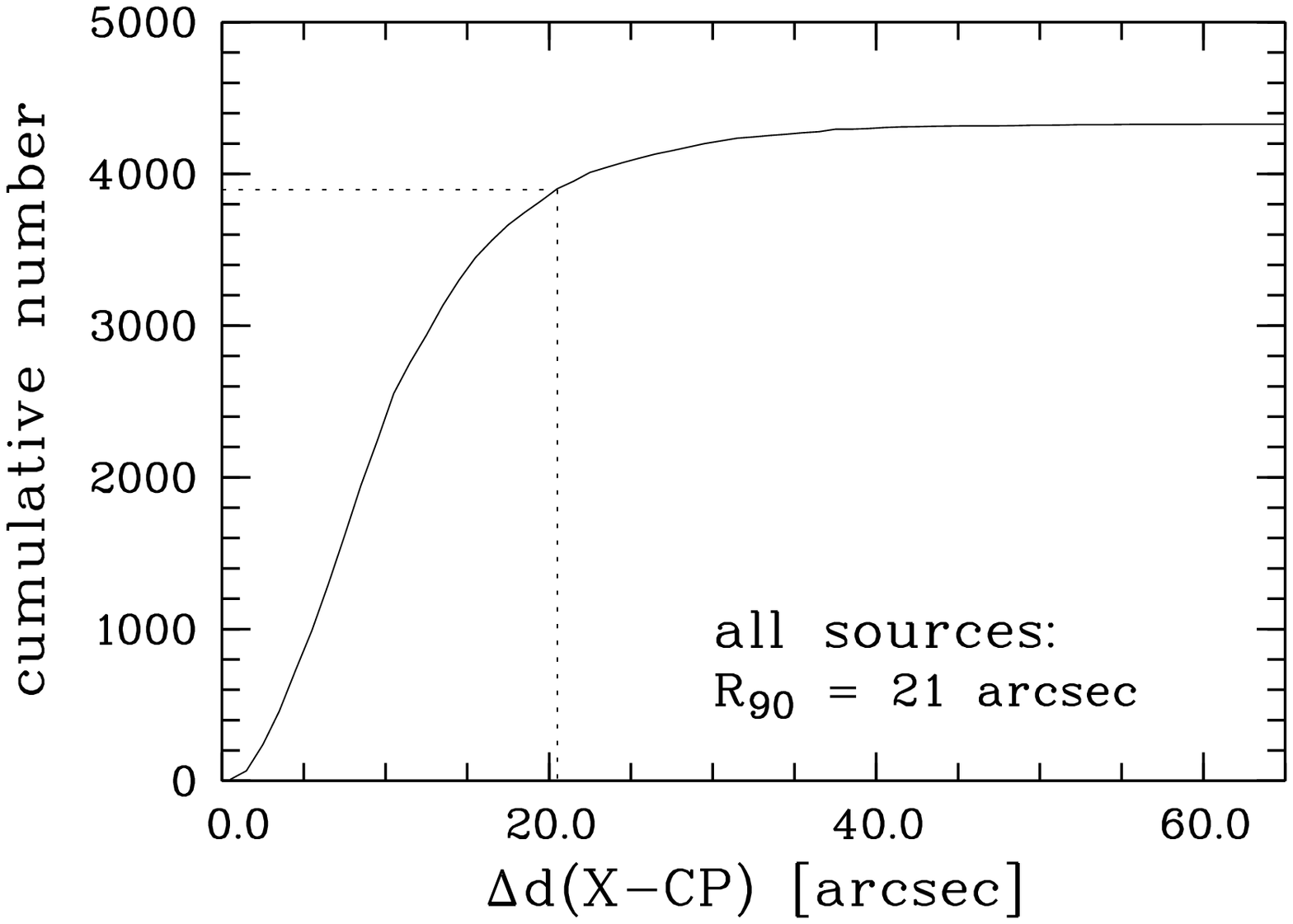}}
\caption[]{Upper panel: difference between R.A. and Dec., respectively, 
of the X-ray and optical position for all point-like counterparts. 
The histograms show the frequency distribution of the positional offsets.
No systematic effects are discernible. Lower panel: cumulative number of sources plotted
over distance between optical counterpart and X-ray position.  90\% of the optical
counterparts are within 21\arcsec\ from the X-ray source.
}
\label{error}
\end{figure}

\begin{figure} 
\resizebox{\hsize}{!}{\includegraphics[angle=-90,bbllx=50pt,bblly=40pt,bburx=520pt,bbury=720pt,clip=]{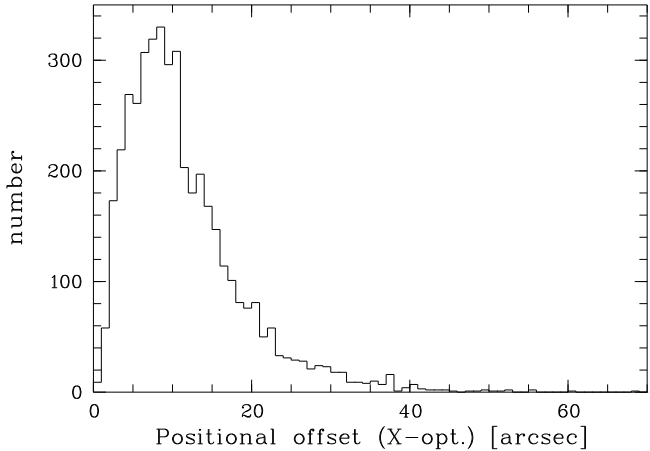}}
\caption[]{Histogram of the positional offsets between X-ray position and position of
the optical counterpart for all X-ray sources identified with optical point sources.
}
\label{disthist}
\end{figure}

\section{Discussion}
\label{discussion}
\subsection{Source statistics}
\label{sourcestat}
In Table \ref{statistics} the statistics of the final identification of the 5341
RASS-BSC sources is summarized. For 82.2\% an identification could be found. 
In 2.9\% of the X-ray sources no optically visible objects were found within
$40\arcsec$ around the X-ray source position (blank fields). Another 14.9\%
remained unidentified, partly due to unusable spectra e.g. because of overlapping spectra, 
but mostly due to optically faint candidates with spectra of too low S/N ratios. 

Experience from follow-up spectroscopy and comparison with the literature shows that
misidentifications are $<$10\%. Therefore, on a case-by-case basis
each of our identifications requires confirmation by optical slit
spectroscopy, and is strictly speaking providing a candidate.
This should be kept in mind in the following discussions, where
we do not explicitly consider  the statistical uncertainty of
our identifications.

The largest group of X-ray emitters
are the active galactic nuclei comprising 41.5\% of all sources. 
Further extragalactic sources are 4.5\% galaxies and 4.9\% clusters of galaxies. 
The second largest group are the bright stars 
with a share of 22.8\%. 
F-G stars only 
contribute 0.8\% due to the fact 
that most of these stars are so bright that their spectra are
saturated and hence fall into the class of bright stars. K and M stars contribute
2.6\% and 3.7\%, respectively. Likewise, small contributions of 0.8\% and 0.5\% are due to 
white dwarfs and cataclysmic variables, respectively.
Hence we identified $\sim$51\% with extragalactic sources and $\sim$30\% with 
stellar coronal emitters. 

These numbers can be compared with the results of other programs for 
optical identification  of X-ray selected samples. 
We first consider two
surveys based like our program on the RASS, i.e. on all-sky observations in the 
soft energy band. These are the LSW-INAOE-MPE survey (LSW survey hereafter) 
(Zickgraf et al. \cite{Zickgrafetal97}, Krautter et al. \cite{Krautteretal99}) 
comprising RASS sources down to count rates of 
0.01\,cts\,s$^{-1}$ and the NEP survey with an even deeper RASS exposure (Henry et al. 
\cite{Henryetal01}), reaching count rates as low as 1.2\,10$^{-3}$\,cts\,s$^{-1}$ (Voges et
al. \cite{Vogesetal01}). Secondly, we compare our results with the EINSTEIN Medium 
Sensitivity Survey (EMSS) which in contrast to the RASS is  
based on serendipitous instead of all-sky observations. Moreover, the energy band of the 
EMSS, 0.3-3.5\,keV,  is harder than that of the RASS. 
Table \ref{comparison} summarizes the comparison. 
In order to account for the contributions to the individual object classes in the HRC 
from unidentifed sources (ID codes 8 and 803) 
and blank fields (ID code 0) we use estimates derived 
below in Sects. \ref{blank} and \ref{unidsource} from the investigation of the hardness 
ratios of the different object classes. The fractions corrected  with these estimates 
are given in Table \ref{comparison} in an additional row for the HRC.

\subsubsection{Comparison with the LSW and NEP surveys}
For the comparison with the LSW survey we exclude area I of this
sample. It has an unusually high column density of neutral hydrogen and a
different source composition compared to the other five areas
(cf. Zickgraf et al. \cite{Zickgrafetal97}). This is probably due to contamination by the 
close Tau-Aur star-forming region located about 20$\degr$ north of area I.

The fraction of coronal emitters in our HRC sample appears smaller than in the reduced 
LSW sample, which contains 36\%  coronal sources compared to our 30\%. The difference is 
at least partly due to the 
difficulty of identifying 
the faint M dwarfs on our blue prism plates. 
Many of these stars appear probably in the RED-WK class 
and are classified here as unidentified sources. 
Note, that M dwarfs in the BSC sample may be as faint as 
$B\sim$17.5 to 18, 
which is not far from the prism plate limit. Taking this into account by omitting the M dwarfs 
from the statistics leads to 26.2\% coronal emitters in the HRC, and 25.8\% in the LSW sample. 
Hence, no significant difference seems to be present  between the HRC and the LSW survey 
for stellar sources with spectral types from F to K. For the NEP Henry 
et al. only give a total of 34\% stars which contains an unknown fraction of M dwarfs. 
Correction for the estimated contribution of stars in the HRC  from unidentified 
sources and blank fields  yields a stellar fraction slightly larger than in the LSW and the NEP 
survey, namely $\sim$38\% stars in the HRC versus 36\% and 34\%,
respectively. The indicated trend of a decreasing stellar fraction with decreasing flux limit 
could be due to the scale height of the stellar density distribution in the Milky Way. 
Note however that the differences are on a 2$\sigma$ level only.

Whereas we found 41.5\% AGN, the reduced LSW sample has an AGN fraction of 46\%.
The NEP survey with its even deeper RASS exposure yielded an even larger fraction 
of 51\% AGN (including BL Lac objects) indicating an increasing AGN fraction with 
decreasing X-ray flux limit. 
Correction for the estimated contribution of AGN in the HRC from unidentified sources 
and blank fields, 
however, yields $\sim$49\% of AGN. This value is between the AGN fractions in 
the LSW and NEP surveys of 46\% and 51\%, respectively.  Within the 2$\sigma$ limits  
of the LSW and the NEP survey no significant difference is found between the three surveys. 

The galaxy fraction in the LSW survey, 4\%, was similar to our result.
However, the fraction of galaxy clusters in the LSW sample is more than a factor 
of two larger (12\%). 
Actually, Krautter et al. (\cite{Krautteretal99}) found the highest fraction of galaxy clusters
in the deepest exposed part of their RASS sample which is  a factor of 5
more sensitive than the BSC sample. Furthermore, the optical CCD $R$ and $B$ images 
used 
by them to identify galaxy clusters reached fainter limiting magnitudes 
(23$^{\rm m}$ and 22$^{\rm m}$ in $R$ and $B$, respectively, cf. Zickgraf et al. 
\cite{Zickgrafetal97}) than our blue plates which only reach as deep as $B \approx 20$.  
This explains the different fraction of galaxy clusters in the
two samples. Likewise, the NEP survey contained 14\% galaxy clusters, again
significantly more than in the HRC, but only 0.2\% galaxies. 
Like the LSW survey the NEP identification was also based on deep red CCD images. 
In addition the deep RASS-NEP exposure allowed to detect extended X-ray sources 
alleviating the identification of galaxy clusters. A part of the missing clusters 
in the HRC could be hidden among the unidentified sources and the  blank fields in particular. As will be discussed in Sects. 
\ref{blank} and \ref{unidsource} we expect in fact that a significant fraction of these  
sources in our BSC sample will be galaxy clusters. Taking the possible 
contributions from these sources to the galaxy cluster fraction into account (see Table 
\ref{comparison}) actually diminishes the differences with the LSW and NEP survey somewhat. 
With corrected 7\% the differences relative to the LSW and NEP surveys are still 
5\% and 7\%, respectively. These
differences are on a 4$\sigma$ level. Note, however that a fraction
of the sources identified as galaxies could actually be galaxy clusters. 
This might e.g. be the case for clusters in which only the cD galaxy is visible on our 
direct plate. Adding the fractions
of galaxies and galaxy clusters, no significant difference of the HRC (13\%), 
the LSW (16\%) and the NEP (15\%) is found. Taking these numbers and the possibility of
misidentification into account we are led to
the conclusion that the cluster fraction in the HRC is also not significantly different 
from the LSW and the NEP.
\subsubsection{Comparison with the EMSS}
Comparison with the  EINSTEIN Medium Sensitivity Survey (EMSS) (Stocke et al.
\cite{Stockeetal91}) reveals small differences in the sample composition. 
The AGN fraction in the EMSS, 54\%, is significantly higher than in the uncorrected HRC
with 42\%. For the corrected AGN fraction in the HRC, though  still $\sim$5\% below the 
EMSS, the difference corresponds to only $\sim2\sigma$.
On the other hand, stellar sources are less frequent in the EMSS (26\%) than in the HRC 
(31\% uncorrected). The corrected stellar fraction of the HRC (38\%) is even $\sim$12\% 
above the EMSS. This difference is significant on a $\sim6\sigma$ level.  

With (uncorrected) 5\% the fraction of galaxy clusters in the HRC would be significantly  
lower than the 12\% in the EMSS.  Note that in the EMSS as in the LSW and 
NEP surveys the galaxy cluster identification  was based on deep red CCD images. 
Comparing with the corrected 7\%  the HRC still seems to be deficient in galaxy 
clusters relative to the EMSS. The difference is $\sim4\sigma$. As discussed above 
there is the possibility of galaxy clusters being identified as galaxies.  
Adding the fractions
of galaxies and galaxy cluster together, no significant difference of the HRC (13\%) 
and the EMSS (15\%) is found. As for the LSW and the NEP we can thus state that 
the cluster fraction in the HRC is not significantly different from the EMSS. 

Considering the number ratio of AGN and
stellar coronae reveals a clear and significant difference between the EMSS and the RASS
samples. The ratios (and their 1$\sigma$ uncertainties) are 2.08(0.03), 1.35(0.01),
1.29(0.01), 1.27(0.04), and 1.50(0.04) for EMSS, HRC, HRC$_C$, LSW, and NEP, respectively.
Hence the number ratio of AGN and stars in the EMSS is
significantly different from the HRC and the other RASS based surveys.

A plausible explanation for this finding is the 
harder X-ray energy band of the EMSS compared to ROSAT. The stellar subsample 
consists mainly of coronal sources emitting predominantly in the soft X-ray band.
They are thus expected to occur with a higher frequency in a 
survey at softer X-ray energies compared to harder energies.  This has already been suggested
by Gioia et al. (\cite{Gioiaetal90}).

For the galaxy clusters the  harder energy band of the
EMSS seems not to play a dominant role for the detection efficiency
although the energy spectra of galaxy clusters are hard. This follows
from the similarity of the cluster fractions in the X-ray hard EMSS and  the X-ray 
soft LSW and NEP surveys (see Table \ref{comparison}). 

Apart from effects due to the the energy bands the different AGN to star ratios could 
also at 
least partly be caused by the selection of the X-ray sources from the IPC pointing of the 
EINSTEIN observatory. The EMSS is not an all-sky survey but consists of serendipitous
X-ray sources in the neighborhood of X-ray bright objects.  
One fourth of the IPC fields used for
the EMSS were pointing towards galactic objects. One third had an AGN as target. 
The targets of the pointings were excluded from the EMSS sample. 
This could introduce a bias in the sample composition. For a detailed discussion of the 
selection criteria of the EMSS sample 
and possible biases caused by the selection cf. Gioia et al. (\cite{Gioiaetal90}). 

Summing up, it may be said that 
the HRC agrees within the errors of the
source statistics with the three other surveys except for the stellar fraction in the 
EMSS. This difference could be due to the harder EMSS energy band, the
serendipitous character of this survey, or a combination of both.

\begin{table}
\caption[]{Statistics of the identifications. ``code'' is the 
identification code used in the catalogue. The corresponding object class is listed in
column ``class''. The absolute number and fraction in percent are given in the last two
columns.}   
\begin{tabular}{llll}
\noalign{\smallskip}
\hline
\noalign{\smallskip}
code & class & number& fraction [\%]\\
\noalign{\smallskip}
\hline
\noalign{\smallskip}
0 & blank field& 155&2.9\\
1 &  AGN  & 2215& 41.5\\
2 & galaxy& 238&4.5\\
3 & galaxy cluster& 262&4.9\\
5 & M star& 197&3.7\\
6 & white dwarf& 45&0.8\\
$7\_1$ & K star  & 141&2.6\\
$7\_2$ & F-G star & 45&0.8\\
$7\_3$ &CV& 26&0.5\\
$7\_4$ & bright stars & 1219&22.8\\
8 & unidentified& 604&11.3\\
803 & no spectrum & 194&3.6\\
\noalign{\smallskip}
\hline
\noalign{\smallskip}
Total: &  &5341 &100\\
\noalign{\smallskip}
\hline
\noalign{\smallskip}
\end{tabular}
\label{statistics}
\end{table} 

\begin{table}
\tabcolsep=5pt
\caption[]{Comparison of the HRC statistics with the LSW-INAOE-MPE (``LSW'' ), 
the NEP survey (``NEP''), and the EMSS. 
For each object class the percentage is given.
Remarks: $^1$including BL Lac objects, $^2$without area I 
(Zickgraf et al. \cite{Zickgrafetal97}), 
$^3$includes 3\% blank fields. 
In row ``HRC$_{\rm C}$'' the unidentified sources and blank fields have been
redistributed into the classes AGN, galaxy cluster, 
galaxies, and stars based on hardness ratio distributions. 
1$\sigma$ errors are given in parentheses.
}   
\begin{tabular}{llllll}
\noalign{\smallskip}
\hline
\noalign{\smallskip}
survey               & AGN$^1$ & gal. & gal. cl.& stars & unid. \\
\noalign{\smallskip}
\hline
\noalign{\smallskip}
HRC                  & 42\,(0.9)  & 5\,(0.3) &5\,(0.3)        &31\,(0.8)     &18$^3$\,(0.6)\\
{\bf HRC$_{\rm C}$}  &{\bf 49}\,(1.0) &{\bf 6}\,(0.3)&{\bf 7}\,(0.4)&{\bf 38}\,(0.8)&{\bf -}\\
LSW$^2$              &46\,(2.6) & 4\,(0.8)& 12\,(1.3)& 36\,(2.3)& 2\,(0.5) \\
NEP                  & 51\,(3.4) & 1\,(0.5)& 14\,(1.8)& 34\,(2.8)&1\,(0.5)\\  
EMSS                 & 54\,(2.5)& 3\,(0.6)&{\bf 12}\,(1.2)& 26\,(1.8)& 4\,(0.7)\\
\noalign{\smallskip}
\hline
\noalign{\smallskip}
\end{tabular}
\label{comparison}
\end{table}

\begin{figure} 
\resizebox{\hsize}{!}{\includegraphics[angle=-90,bbllx=40pt,bblly=40pt,bburx=530pt,bbury=730pt,clip=]{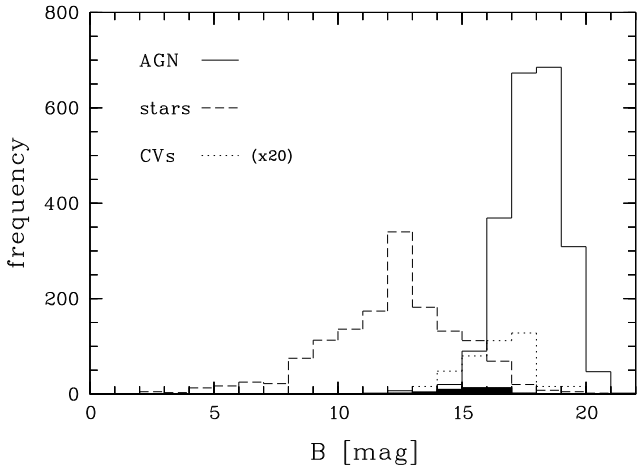}}
\caption[]{Histograms showing the distribution of B magnitudes for AGN, stars  
(bright stars, F-G, K, M stars, white dwarfs), and CVs.
For the latter the ordinate scale has been expanded by a factor of 20  
to account for the smaller numbers. The contribution of white dwarfs is additionally 
plotted as black filled histogram.
}
\label{bmag}
\end{figure}

\begin{figure} 
\resizebox{\hsize}{!}{\includegraphics[angle=-90,bbllx=40pt,bblly=40pt,bburx=530pt,bbury=730pt,clip=]{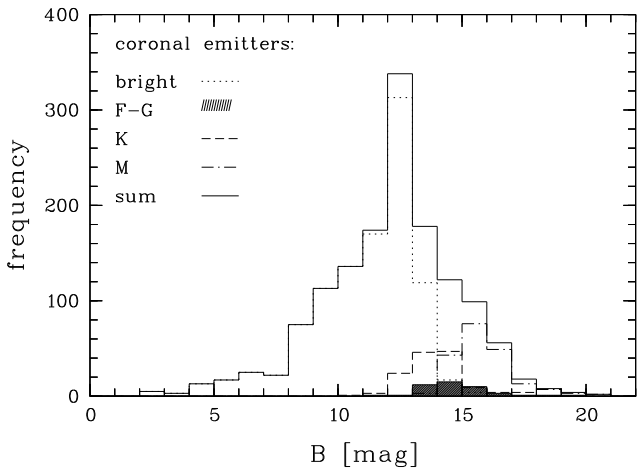}}
\caption[]{Histograms showing the distribution of B magnitudes for coronal emitters
i.e. bright stars, F-G, K, and M stars. The small 
contribution from F-G stars is shown as shaded histogram.
}
\label{bstars}
\end{figure}

\subsection{Positional accuracy}
In Fig. \ref{error} the positional offsets of the optical counterpart and the X-ray
position are plotted for right ascension and declination. No significant systematic 
shift is discernible in either coordinate. The lower panel of the figure depicts the
cumulative distribution of distances between optical and X-ray position. 
We found that 90\% of the point-like
counterparts are found within 21$\arcsec$ from the RASS position, i.e.
the 90\% error radius $r_{90} = 21\arcsec$. This is significantly smaller than the 
90\% error radius determined from the LSW survey for which Krautter et al. 
(\cite{Krautteretal99}) give 30$\arcsec$. The difference can be understood in terms 
of the better photon statistics for the BSC sources compared to the low count numbers
of the RASS sources in the LSW sample. The frequency
distribution of distances is displayed in Fig. \ref{disthist} showing that the most
frequent distance is $\sim9\arcsec$ with a broad wing towards larger separations.

\subsection{Optical brightness distribution}
\label{optbright}
The distribution of the $B$ magnitudes 
of the counterparts
is displayed in Fig. \ref{bmag}. We have excluded
galaxies because of the unreliable photometry of the optically extended objects.
The brightness distribution of the AGN class comprising Sy galaxies, QSOs and BL Lacs 
spans the range $B\sim14-20$ and peaks around $B \sim17$. The stellar sources,
excluding CVs, have a broader distribution from $B\sim4$ to $B\sim18$ with a 
gradual increase
and a maximum appearing around $B\sim12-13$. The subgroup of white dwarfs 
(black histogram in 
Fig. \ref{bmag}) peaks around $B\sim15$.  CVs start around $B\sim13$ and 
are found up to $B\sim19$ with a peak at $B\sim16$. 

In Fig. \ref{bstars} the brightness distribution 
for coronal emitters is broken down into the different contributions by F-G, K, M, and
bright stars. 
Beyond the  pronounced maximum the number of stars 
decreases in a broad wing towards fainter magnitudes.  

\begin{figure} 
\resizebox{\hsize}{!}{\includegraphics[angle=-90,bbllx=40pt,bblly=40pt,bburx=540pt,bbury=780pt,clip=]{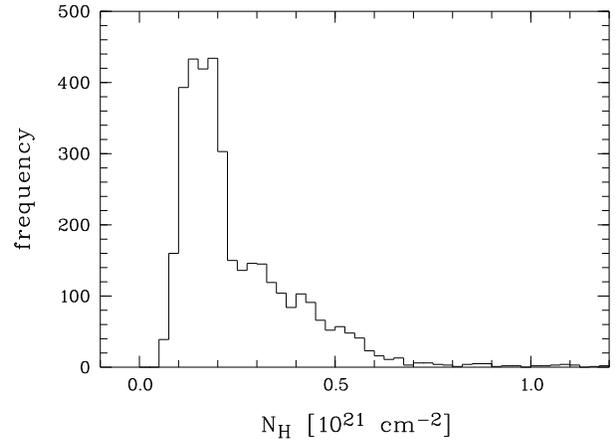}}
\caption[]{Column density distribution of neutral hydrogen of extragalactic X-ray sources. 
}
\label{nh}
\end{figure}

\begin{figure} 
\resizebox{\hsize}{!}{\includegraphics[angle=-90,bbllx=40pt,bblly=40pt,bburx=530pt,bbury=730pt,clip=]{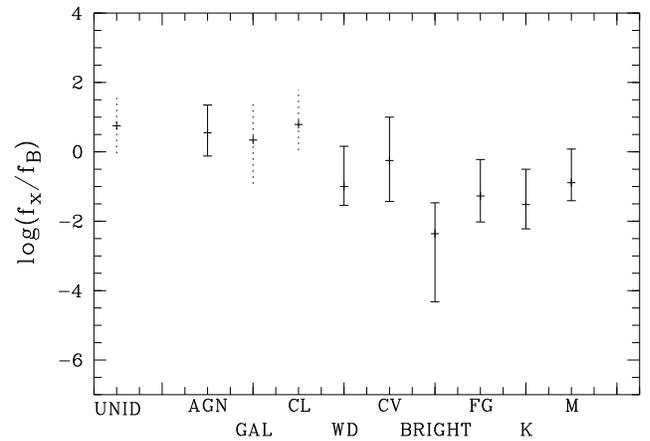}}
\caption[]{X-ray-to-optical flux ratios for the different object classes. The diagram
shows the range of $\log f_{\rm x}/f_B$ containing 90\% of the counterparts and the 
median value of the respective class.
} 
\label{fxfb}
\end{figure}

\subsection{X-ray-to-optical flux ratio}
\label{fxfopt}
In order to determine the ratio of X-ray to optical flux for the various object classes we
calculated the flux in the 0.1-2.4\,keV energy range from the BSC count rates.  
Conversion factors were determined with the WebPimms tool (version v3.2d) 
provided by HEASARC (Mukai \cite{Mukai93}).
For extragalactic sources
column densities of neutral hydrogen, \nh , were obtained from
Dickey \& Lockman (\cite{DickeyLockman90}) using EXSAS (Zimmermann et al. 
\cite{exsas}). The column density distribution is displayed in
Fig. \ref{nh}. It peaks around 1.5\,10$^{20}$\,cm$^{-2}$ with a wing up to 
$\sim1.0\,10^{21}$\,cm$^{-2}$. The median of the distribution is
2\,10$^{20}$\,cm$^{-2}$ and the mean $<$\nh$>$ = 2.6\,$10^{20}$\,cm$^{-2}$. 
For AGN we used a power law photon index of $\Gamma = 2.5$ (see below). For 
individual galaxies and clusters of galaxies  thermal bremsstrahlung with a 
temperature of $T = 1$\,keV and $T = 8$\,keV, respectively, was adopted.
For stellar sources we adopted \nh = 10$^{18}$\,cm$^{-2}$. Except for white dwarfs 
a Raymond-Smith model with $T = 1$\,keV was used. The conversion factor for 
white dwarfs was determined for a black body energy distribution with 
$T = 50\,000$\,K. 

The optical fluxes in the $B$ band were obtained with the flux calibration for 
Johnson $B$ of $F_B(0.0$\mgn) = $7.2\,10^{-9}$\,\flux\ and a filter width of 
$\Delta B = 1000$\,\AA\
(Lamla \cite{Lamla82}). 
This results in the relation 
$\log f_{\rm x}/f_B = \log f_{\rm x} +B/2.5 +5.14$. The
conversion to the usually used ratio $\log f_{\rm x}/f_V$ is given by  
$\log f_{\rm x}/f_B - \log f_{\rm x}/f_V = (B-V)/2.5 - 0.36$ with $F_V(0.0$\mgn)$ = 
3.92\,10^{-9}$\,\flux\ and $\Delta V = 800$\,\AA . Hence differences are small,
-0.36 for $B-V = 0.0$ and +0.24 for $B-V = +1.5$.

The resulting flux ratios  $\log f_{\rm x}/f_B$ are listed in Table \ref{fxfbtab} 
and displayed in Fig. \ref{fxfb}. 
For each object class the median and the limits of the frequency distributions 
at the 5\% level, i.e. the range containing 90\% of the sources, are given. 
For unidentified sources only in a few cases 
$B$ magnitudes of spectroscopically unidentified but likely 
optical counterparts could be listed in the catalogue with some confidence. 
Magnitudes of galaxies are uncertain due to the extended images.
Likewise, only for a few galaxy clusters approximate $B$ magnitudes for the 
brightest members in the error circle could be used to obtain flux ratios. 
The ranges for these classes are therefore uncertain and are plotted as dotted 
lines. The flux ratios for galaxies and galaxy clusters are less reliable due to the
selection effects and uncertain photometry for galaxies.

\begin{table}
\caption[]{Flux ratios  $\log f_{\rm x}/f_B$ for the various object classes. 
Column 2 lists the median value and column 3 the range containing 90\% of the sources.
}   
\begin{tabular}{llll}
\noalign{\smallskip}
\hline
\noalign{\smallskip}
class            & $<\log f_{\rm x}/f_B>$ & \multicolumn{2}{c}{range} \\
                 &                        & \multicolumn{1}{c}{low} & \multicolumn{1}{c}{high}  \\
\noalign{\smallskip}
\hline
\noalign{\smallskip}
AGN              & +0.55   & $-0.12$  &   +1.35\\
galaxies         & +0.34:  & $-0.92$: &  $+1.45$: \\
galaxy clusters  & +0.79:  & $+0.05$: &  $+1.78$:\\
bright stars     & $-2.36$ & $-4.32$  &  $-1.47$\\
F-G stars        & $-1.27$ & $-2.02$  &  $-0.22$\\
K stars          & $-1.52$ & $-2.22$  &  $-0.50$\\
M stars          & $-0.89$ & $-1.41$  &  $+0.08$\\
white dwarfs     & $-1.00$ & $-1.54$  &  $+0.16$\\
CVs              & $-0.25$ & $-1.43$  &  $+1.00$\\
unidentified     & +0.75:  & $-0.04$: &  $+1.64$:\\
\noalign{\smallskip}
\hline
\noalign{\smallskip}
\end{tabular}
\label{fxfbtab}
\end{table}

\begin{figure*} 
\centering
\includegraphics[width=17cm,angle=0,bbllx=75pt,bblly=120pt,bburx=580pt,bbury=610pt,clip=]{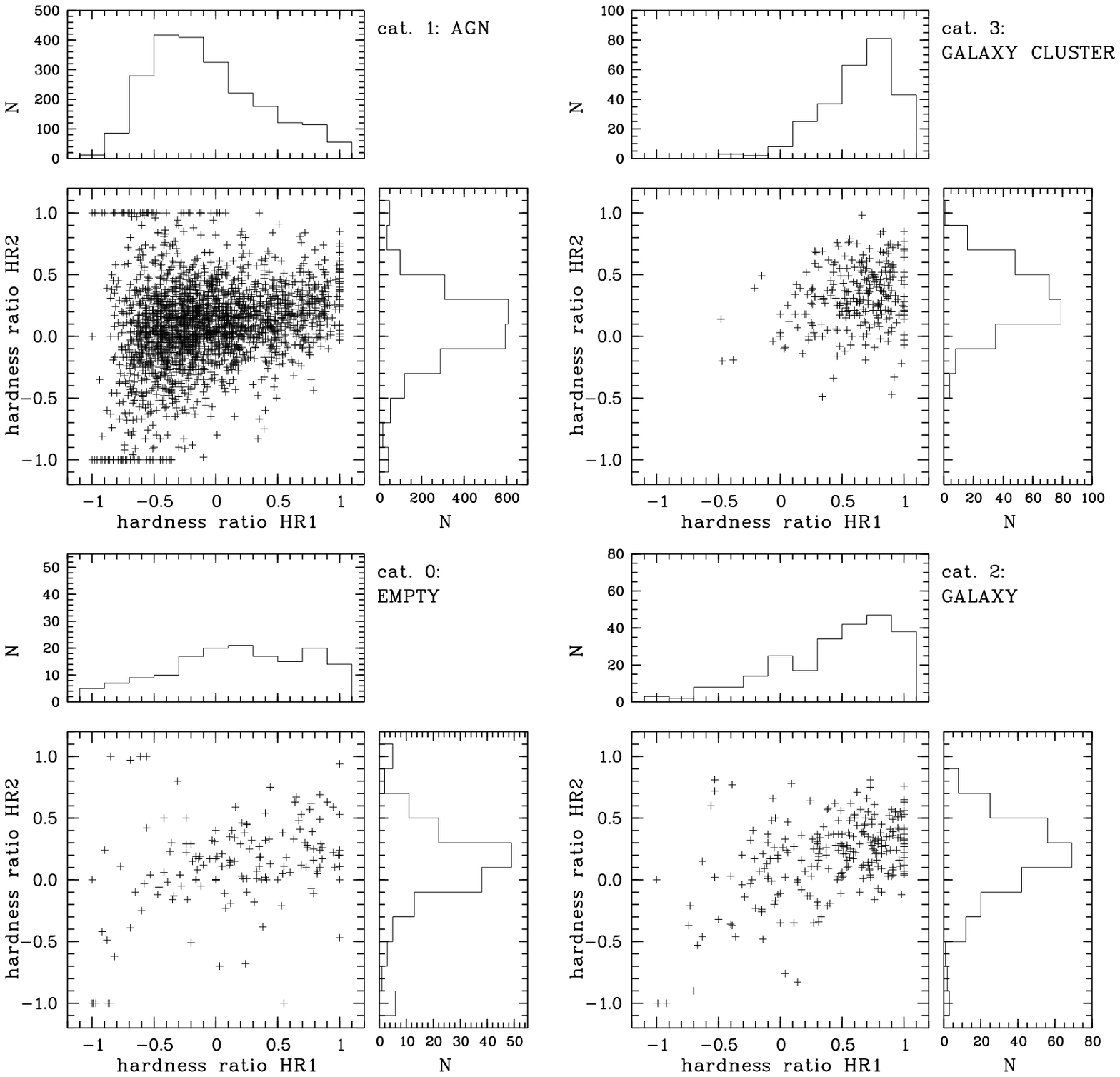}
\caption[]{Hardness ratios of extragalactic sources and blank fields ("EMPTY").
}
\label{hardexgal}
\end{figure*}

The median $\log f_{\rm x}/f_B$ for AGN is {\bf +0.55}. Accounting for the colour
correction (+0.16 for $B-V =0.5$) this is in very good agreement with the 
result of Krautter et al. (\cite{Krautteretal99}) of $<\log f_{\rm x}/f_V> = +0.44$
for the LSW sample. 
Among the stellar counterparts CVs have the highest flux ratio, 
$<\log f_{\rm x}/f_B> = -0.25$. Their flux ratio 
range overlaps partly that of AGN. Krautter et al. found a  flux ratio of
$<\log f_{\rm x}/f_V> = +0.42$ for the class of emission-line stars which contains
mainly CVs. This is consistent with our range of flux ratios. 
White dwarfs are about on the same level as M dwarfs which among the 
coronal emitters exhibit the highest flux ratio.
Fig. \ref{fxfb} suggests a trend of increasing $<\log f_{\rm x}/f_B>$ from bright
stars, F-G stars  and K stars to M dwarfs. Note, however that the group of 
bright stars contains mainly F-G
stars and should therefore be combined with these. The combined flux ratios of the
two groups thus span the range 
from $-4.32$ to $-0.22$.

\subsection{Hardness ratios}
\label{defhard}
\begin{figure*} 
\includegraphics[width=17cm,angle=0,bbllx=75pt,bblly=120pt,bburx=580pt,bbury=610pt,clip=]{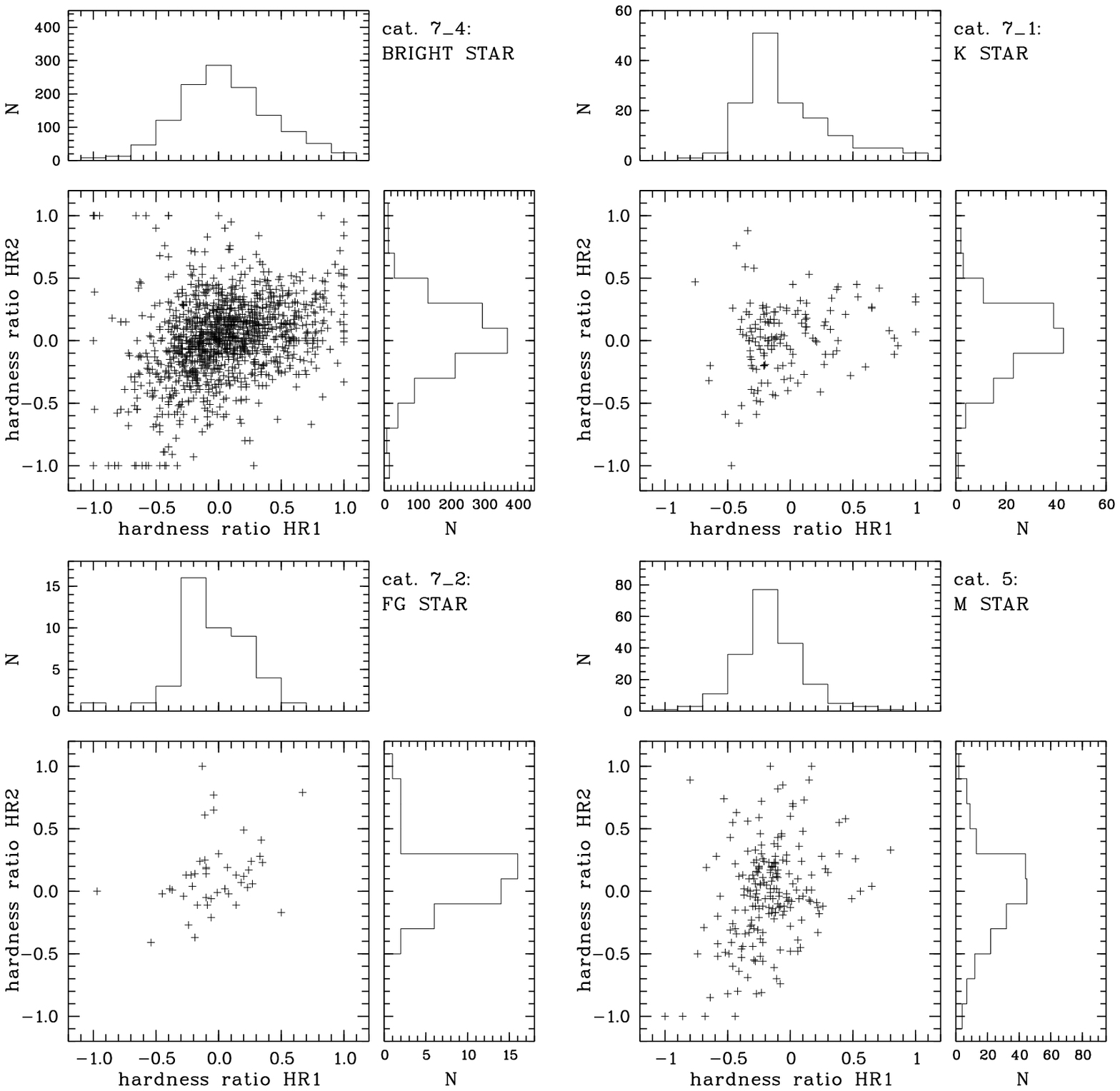}
\caption[]{Hardness ratios of bright stars, F-G, K, and M stars. 
}
\label{hardstar}
\end{figure*}

\begin{figure} 
\resizebox{\hsize}{!}{\includegraphics[angle=0,bbllx=200pt,bblly=120pt,bburx=445pt,bbury=605pt,clip=]{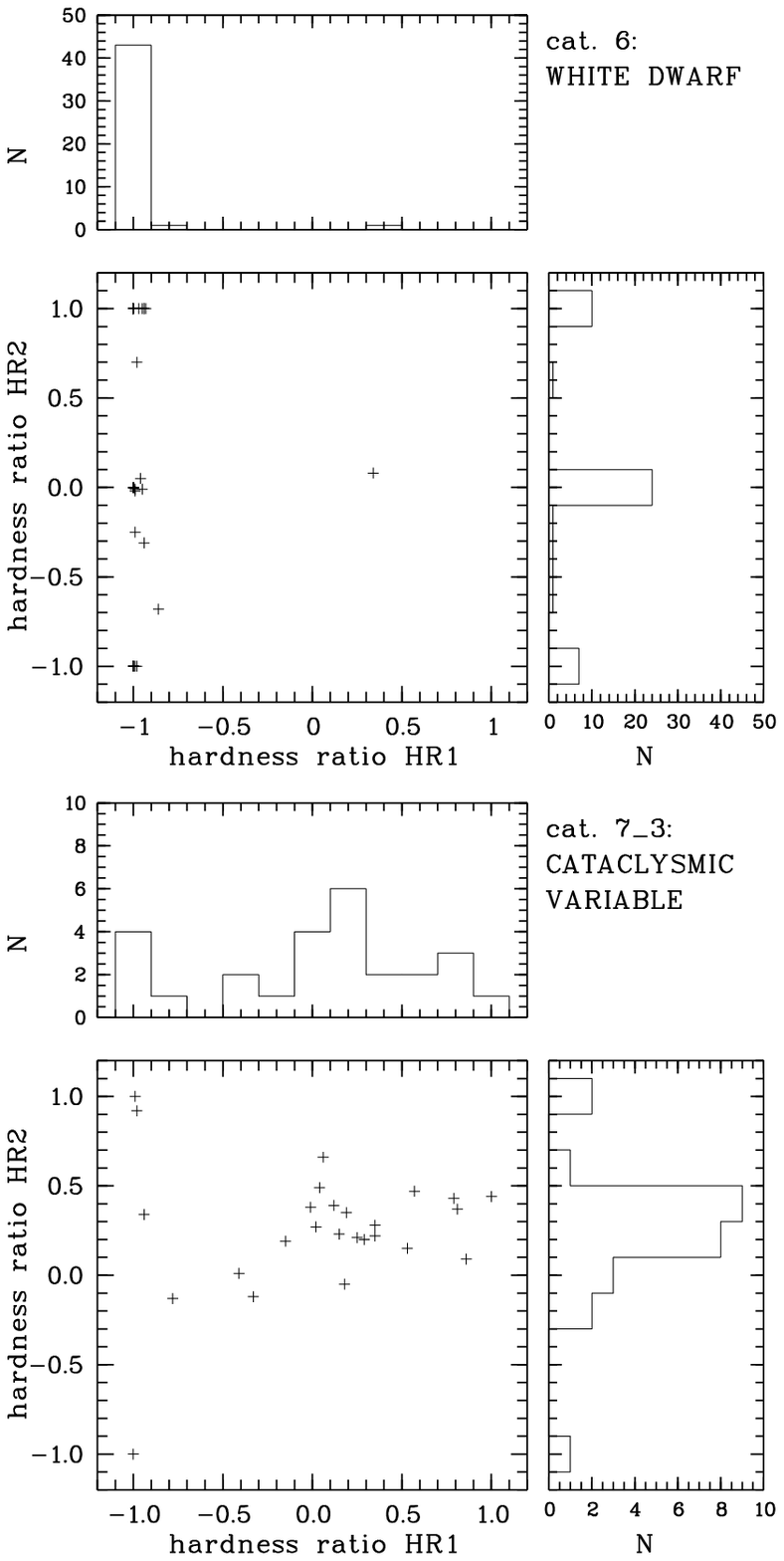}}
\caption[]{Hardness ratios of white dwarfs and cataclysmic variables. 
}
\label{hardcompact}
\end{figure}

\begin{figure} 
\resizebox{\hsize}{!}{\includegraphics[angle=-90,width=8.8cm,bbllx=40pt,bblly=45pt,bburx=560pt,bbury=580pt,clip=]{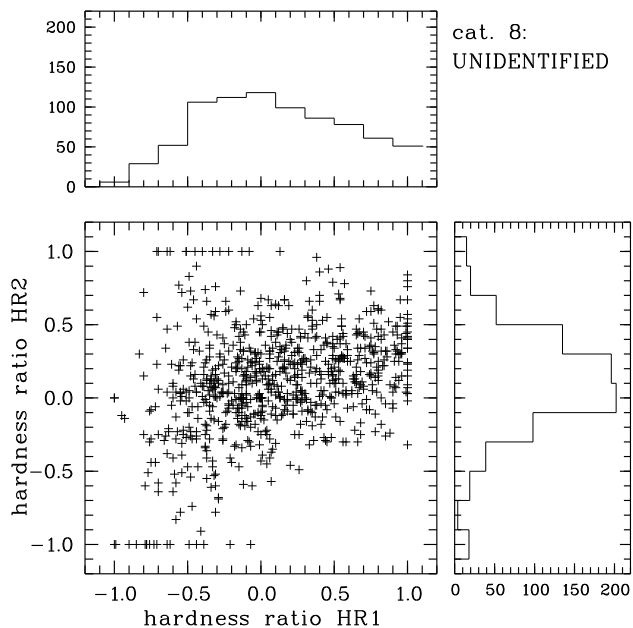}}
\caption[]{Hardness ratios of unidentified sources. 
}
\label{hardunid}
\end{figure}

In Figs. \ref{hardexgal}, \ref{hardstar},  \ref{hardcompact}, and \ref{hardunid} the hardness ratio $HR1$ is  plotted
versus  $HR2$ for the different classes. Hardness ratios are defined as 

\begin{displaymath} 
HR1 = \frac{[B]-[A]}{[B]+[A]} 
\end{displaymath}
and 
\begin{displaymath} 
HR2 = \frac{[D]-[C]}{[D]+[C]},
\end{displaymath}
where $[A]$ to $[D]$ are the count rates in the respective energy bands 
$A = 0.11-0.41$\,keV, $B = 0.52-2.01$\,keV, $C = 0.52-0.90$\,keV, and 
$D = 0.91-2.01$\,keV. 

\subsubsection{Extragalactic sources}
For AGN the distribution of $HR1$ is broad
with a maximum around $-0.4$ whereas the distribution of $HR2$ is narrower and peaks
at $HR2=$0.0. Galaxy clusters show significantly harder spectra with $HR1$ having a
maximum at $+0.8$. Likewise $HR2$ is harder than for AGN, although less 
conspicuous, with a peak at $+0.2$. The hardness ratio distribution of galaxies 
resemble that of galaxy cluster when comparing the maximum of $HR1$, although the
distribution of the galaxies has a flatter shape at smaller $HR1$. 
This indicates that possibly a fraction of the hard sources
identified with individual galaxies in reality are galaxy clusters. 

\subsubsection{Stellar sources}
Considering bright stars and F-G stars as one class the bulk of these objects 
has hardness ratios around 0 for both, $HR1$ and $HR2$.  K and M stars appear 
to have slightly
softer $HR1$ with a maximum around $-0.2$. Cataclysmic variables have a broad
distribution of $HR1$ with more hard than soft sources and a peak around 
$HR1 \approx +0.2$ whereas $HR2$ has a narrower peak at $+0.4$. All 
white dwarfs but one exhibit very soft $HR1$ of $-1$. The object
with $HR1 = +0.4$ (RX\,J1523.1+3053) though showing the optical 
spectrum of a white dwarf could be a misidentification because an optically faint 
background galaxy cluster cannot be excluded.

\subsubsection{Blank fields}
\label{blank}
\begin{figure} 
\resizebox{\hsize}{!}{\includegraphics[angle=-90,bbllx=45pt,bblly=35pt,bburx=550pt,bbury=745pt,clip=]{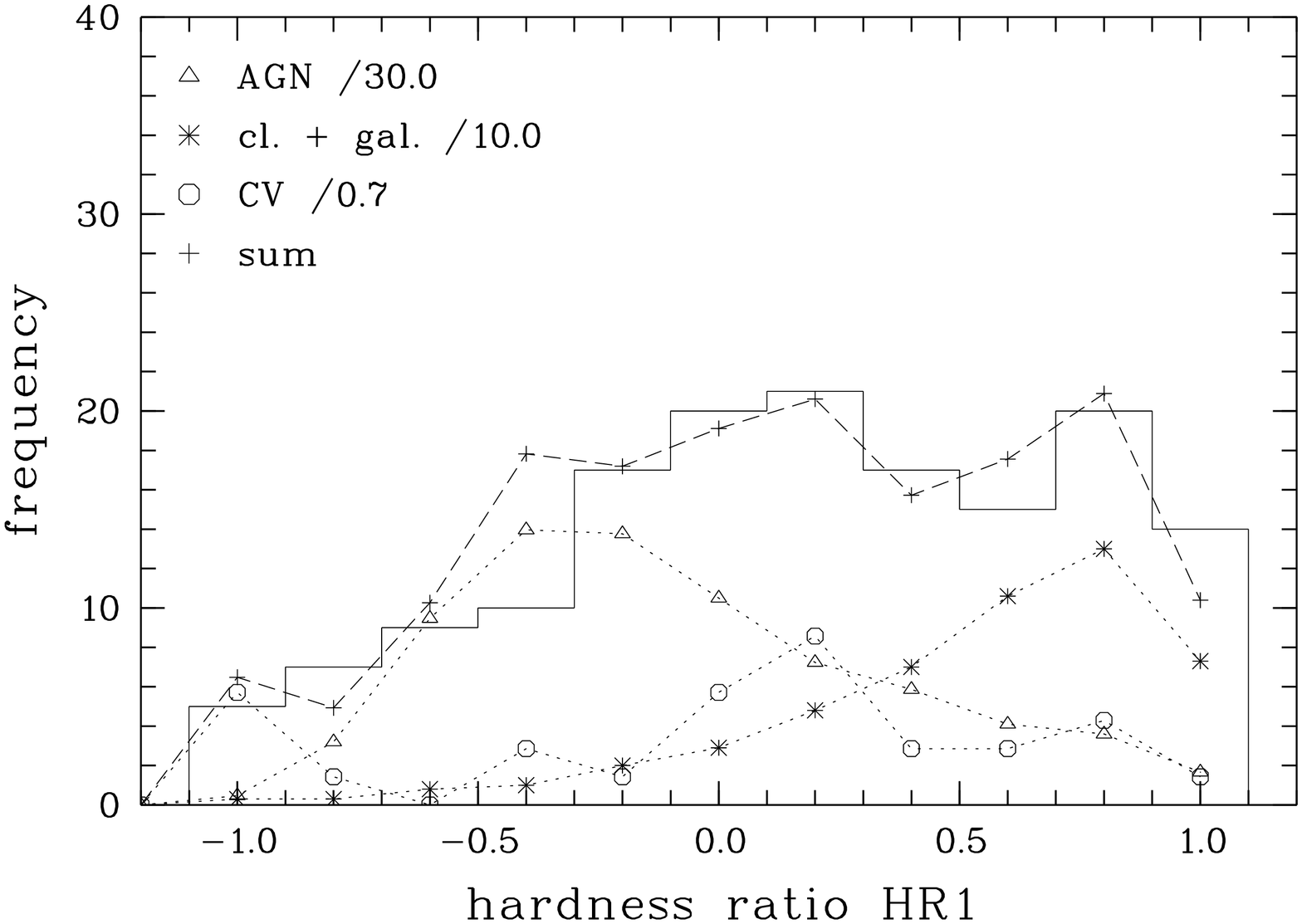}}
\caption[]{Histogram of the hardness ratio $HR1$ of blank fields  (solid line). The dashed lines
represent scaled frequency distributions of AGN, CVs, and galaxy clusters. The
scaling factors  given were
chosen so that the sum of the scaled distributions
(long dashed lines and + signs) approximates the observed histogram. 
}
\label{hardmod}
\end{figure}

The distribution of $HR1$ of the blank fields has a shallow maximum around $+0.2$ 
and possibly a
second peak near $+0.8$. 
Likely counterparts to blank fields are optically faint galaxy clusters, 
individual galaxies, AGN and CVs. Given the X-ray flux limit and the upper limits of 
the X-ray-to-optical flux ratios of the various classes 
of X-ray emitters counterparts of other object classes would be visible 
on our blue plates at a limiting magnitude of $B_{\rm lim} = 20$. 
The overall shape of the blank field histogram of $HR1$ can be
approximated by scaling and adding the contributions of the 
four classes. 
This is shown in Fig. \ref{hardmod}. 
For galaxy clusters and galaxies the combined histogram was used.
The decomposition performed in this way suggests that 
about 30\% of the blank fields are galaxies or galaxy clusters, 
45\% are AGN, and 25\% are CVs.  
The high CV contribution seems necessary in order to explain the broad peak around $HR1
\approx +0.2$ which cannot be reproduced with AGN,  
galaxies, 
and galaxy clusters alone. It would thus
imply that our sample contains possibly more than twice as much CVs as listed in Table
\ref{statistics}.

\subsubsection{Unidentified sources}
\label{unidsource}
\begin{figure} 
\resizebox{\hsize}{!}{\includegraphics[angle=-90,bbllx=45pt,bblly=35pt,bburx=550pt,bbury=745pt,clip=]{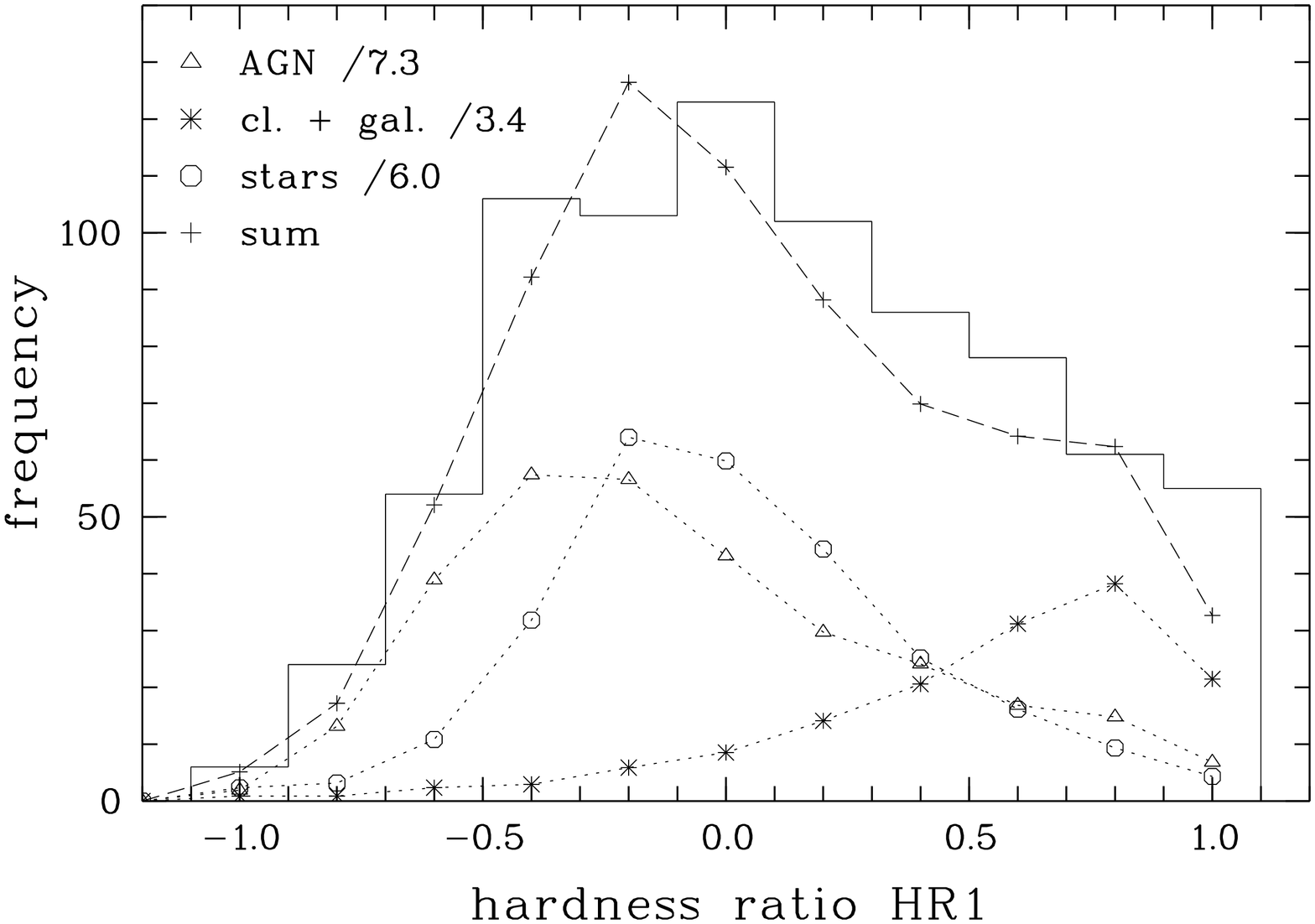}}
\caption[]{Histogram of the hardness ratio $HR1$ of unidentified sources  (solid line). 
The dashed lines represent scaled frequency distributions of AGN, stars, and 
galaxy clusters. The scaling factors  given were
chosen so that the sum of the scaled distributions
(long dashed lines and + signs) approximates the observed histogram. 
}
\label{hardunidmod}
\end{figure}

The distribution of $HR1$ for unidentified sources 
is broad and asymmetric
with a steep decrease between $HR1 \approx -0.4$ and $-0.8$, 
a maximum around $HR1 \approx 0.0$ and with more hard  ($HR1 \ge 0.0$) 
than soft sources ($HR1 < 0.0$). 
We decomposed the distribution in a similar way as
the blank fields.  Galaxies and galaxy clusters were  combined as before. 
Stellar counterparts were all summed up in one histogram for  
``stars''. The third class contributing to the unidentified sources are AGN.
The result of the decomposition is depicted in Fig. \ref{hardunidmod}.
Stars and AGN seem to contribute equal shares of about 40\% to the unidentified sources  
whereas only a fraction of 20\% is probably due to optically faint galaxy clusters and
galaxies.

Taking the galaxy cluster and galaxy contributions from both, blank fields and unidentified 
sources, into account would increase the total fraction of galaxy clusters from 
4.9\% to {$\sim7$\%} and the galaxies from 4.5\% to $\sim6$\%. For these estimates the
combined numbers of galaxy clusters and galaxies were divided according to their relative
fraction among the identified sources.
Likewise, the AGN fraction would increase to 49\% and the stellar fraction to 38\%  
(cf. Table \ref{comparison}). 

\subsection{Continuum properties of AGN}
\subsubsection{Radio emission}
A basic property of AGN is their radio emission which is believed to be related 
to the presence of 
more or less prominent radio jets emerging from the central engine. Based on the ratio of 
radio to optical flux they are separated 
in radio-loud and radio-quiet AGN. Recent studies of radio-loud X-ray selected 
AGN  have been carried out e.g. by Laurent-Muehleisen et al. (\cite{Laurent97}), 
Perlman et al. (\cite{Perlmanetal98}), and Landt et al. (\cite{Landtetal01}). 

In their review on radio-loud AGN Urry \& Padovani (\cite{UrryPadovani95})
quote a fraction of 15-20\% of radio-loud AGN.  
This fraction increases with optical and X-ray
luminosity (Padovani \cite{Padovani93}, La Franca et al. 
\cite{LaFrancaetal94}, Della Ceca et al. \cite{DellaCecaetal94}) to up to 50\%.
Due to the lack of redshifts for most AGN from our HRC sample we cannot 
investigate the dependence of the radio-loud to radio-quiet fractions on luminosity. 
We can, however, obtain at least a general picture of how both classes are represented 
in the HRC and how this compares with AGN samples selected by different methods.

\begin{figure} 
\resizebox{\hsize}{!}{\includegraphics[angle=0,bbllx=45pt,bblly=45pt,bburx=545pt,bbury=745pt,clip=]{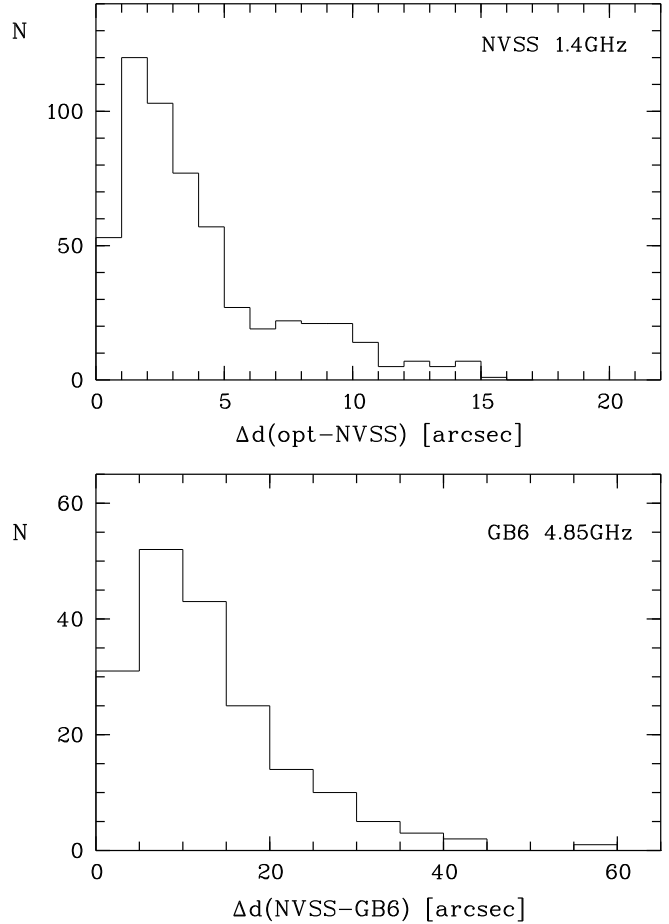}}
\caption[]{Histograms showing the frequency of radio sources with given distances between 
optical counterpart and matching NVSS (upper panel) and between NVSS and GB6 radio sources
(lower panel), respectively.}
\label{distrad}
\end{figure}

In order to obtain radio fluxes for our AGN sample we 
searched 
in several steps
the NRAO-VLA 1.4\,GHz (NVSS, Condon et al. \cite{Condonetal98}) and the 
Green Bank GB6 4.85\,GHz (Gregory et al. \cite{Gregoryetal96}) catalogues for radio 
sources positionally matching the optical counterparts identified as AGN.  
The radio flux limits of these catalogues are 2\,mJy and 18\,mJy, 
respectively. The search radii were 
initially  
set to 15\arcsec\ for NVSS and 80\arcsec\ for GB6. 
For the NVSS the number of matches was 559 of which 90\% were found within 10\arcsec . 
For the GB6 330 sources were detected with a 90\% radius of 58\arcsec .
Fig. \ref{distrad} depicts the histo\-gram of the number of radio 
sources versus distance bet\-ween NVSS radio and optical position (upper panel). 
In a second step we matched the NVSS and GB6 sources by searching
coincidences within 40\arcsec\ of the NVSS sources. We found 183 positional matches. The
histogram of distances is also shown in Fig. \ref{distrad}. Note,
however, that the declination ranges covered by
the two radio surveys are different. GB6 is limited to $0\degr \le  \delta \le +75\degr$ 
whereas NVSS
contains sources within $-40\degr \le \delta \le +90\degr$. For the coincidences 
we calculated the
spectral index $\alpha_{\rm r} = 1.85\,\log(f_{\rm 4.85\,GHz}/f_{\rm 1.4\,GHz})$, 
where $\alpha_{\rm r}$ is defined
by $f_{\nu} \propto \nu^{\alpha_{\rm r}}$. The frequency distribution of  $\alpha_{\rm r}$ 
has a maximum around $+0.1$ and covers a range 
from $-1$ to $+1$.

The result 
of the the first step of the cross-correlation with 
the NVSS agrees well with that obtained by Bischoff \& Becker
(\cite{BischoffBecker97}) who searched for matches between the NVSS and known quasars. 
They give a 90\% confidence error circle of $\sim$10\arcsec\ for a radio source 
at the flux limit of the survey and conclude that matches beyond 35\arcsec\ are almost
totally chance coincidences. 
While this conclusion may be correct for core-dominated radio sources it
may not hold for lobe-dominated sources. 
For these a larger search radius is appropriate. Our search for coincidences 
could thus  miss lobe-dominated 
sources with core fluxes below the survey limit. In the DXRB survey Perlman et al. 
(\cite{Perlmanetal98}) used a search radius of 90\arcsec\ in order to detect extended 
radio structure. Likewise, in the REX survey Caccianiga et al. (\cite{Caccianigaetal99})
used a large search box for the same reason. In order to detect extended radio sources 
also in our sample 
we performed a further search step with a larger search radius of 90\arcsec . 
For 812 RASS AGN candidates 1 to 3 NVSS sources were found within this radius. 
Of these 101 are double and 14 are triple radio detections. 
Among the double and triple sources 63 and 11, respectively, have one of the radio sources within 
15\arcsec\ from the optical position. They were therefore already contained in the 
sample of 559  sources. The remaining 41 double and triple sources are new. The rest of 212
single sources has a separation larger than
15\arcsec\ between the optical and the radio source. Because their association with the RASS
sources is questionable they were considered as chance 
coincidences.

The 41  new double and triple sources could be cases of lobe-dominated radio sources 
with a core flux below the survey limit. In order to test whether these radio sources
are in fact related to the respective optical counterpart we checked the angles between the
optical and the radio positions (cf. also Caccianiga et al. \cite{Caccianigaetal99}).
Double lobe sources should have an angle 
difference of 180\degr . As test criterion we used the angle difference, allowing for an
uncertainty of  $\pm25\degr$. Certainly, a more detailed analysis would require 
to check the NVSS radio images individually. However, this is beyond the scope of this paper, 
and we restrict the discussion to the use of the angle difference. With this criterion 
the 11 triple sources with  core emission qualify as double-lobe sources. Of the
remaining 3 triple sources 2 are probably  double-lobe sources without core emission. In
these cases the third radio source seems to be unrelated to the optical object. 
One triple source is most likely not related to the optical counterpart of the X-ray source.
For 11 double sources neither the angles difference nor the positional separation of the 
radio and the optical position indicates a relation between the radio sources and the 
optical object. For the remaining 27 cases the separations are between 168\degr\ and 
203\degr , in 22 cases even between 170\degr\ and 190\degr . Based on the angle criterion 
the 28 sources can thus be classified as probable double-lobe sources without detected core 
emission. In summary, 29 of the 41 multiple 
sources can be assigned to the optical counterparts, and 12 are spurious detections. 
This means that the first search step with a 15\arcsec\ search radius missed 
$\sim5\%$ of radio emitting AGN. In total, for 588 AGN radio emission was found. The results of
the cross-correlation are summarized in Table \ref{nvssmatch}.

\begin{table*}
\caption[]{Results of the cross-correlation of the RASS AGN candidates and the NVSS. The type of
NVSS source, distance between optical and radio position, $d({\rm opt-radio})$, and 
the number $N$ of
respective sources are given. 
}   
\begin{tabular}{llll}
\noalign{\smallskip}
\hline
\noalign{\smallskip}
 type of NVSS source           & $d({\rm opt-radio}) $& $N$ & comment  \\
\noalign{\smallskip}
\hline
\noalign{\smallskip}
single             & $d \le 15\arcsec$  &  485 & core-dominated AGN\\
($N = 697$)         & $15\arcsec < d \le 90\arcsec$ & 212 &  association questionable\\
double             & 1 component $d<15\arcsec$  &  63 & core-lobe  AGN\\
($N = 101$)         & aligned  &  27 & lobe-dominated  AGN\\
                   & not aligned  &  11 & association questionable\\
triple             & 1 component $d<15\arcsec$  &  11 & core-lobe  AGN\\
($N = 14$)          & aligned  &  2  & lobe-dominated  AGN\\
                   & not aligned  &  1  & association questionable\\
\noalign{\smallskip}
\hline
\noalign{\smallskip}
\end{tabular}
\label{nvssmatch}
\end{table*} 

In order to determine the radio-loudness we extra\-polated 
the 1.4\,GHz fluxes to 8.4\,GHz and computed the ratio $R = f_{\rm r}/f_{\rm o}$ 
with the  8.4\,GHz radio flux $f_{\rm r}$ and the monochromatic optical flux $f_{\rm o}$.
The choice of 8.4\,GHz was made because some previous studies of the radio-loudness 
of quasars used this radio frequency to determine $R$, e.g. Hooper 
et al. (\cite{Hooperetal96}) for the Large Bright Quasar Survey (LBQS).  
The total radio flux of the multiple sources was determined by  summing up the 
radio fluxes of the individual components.
For sources without GB6 detection 
within 40\arcsec\ of the 
NVSS core source  
$\alpha_{\rm r} = -0.5$ was used for the 
extrapolation. Note that the results of the analysis of the
radio-loudness do not depend significantly on the chosen frequency and  hold as well for
1.4\,GHz and  4.85\,GHz radio fluxes. 
The optical flux, $f_{\rm o}$, for 
$\lambda=2500$\,\AA\ and magnitude $B$ 
was computed from the relation $f_{2500} = -22.55 -0.4\,B$ by Schmidt (\cite{Schmidt68}). 
No K-korrection was performed because the redshifts are unknown.

Radio-loud sources have $R > 10$ (Kellermann et al. \cite{Kellermannetal89}).
For AGN not detected in the NVSS nor found in GB6 an upper limit  $R_{\rm lim}$ was 
calculated by assuming an upper limit of 2\,mJy for the 1.4\,GHz radio flux. 
Radio-quiet AGN have either $R < 10$ if radio flux is detected or $R_{\rm lim} < 10$ 
otherwise.
AGN not detected in the radio surveys are therefore only included in the radio-quiet sample if
their $R_{\rm lim} < 10$. For the remaining radio non-detections an assignment to eiter the
radio-quiet or radio-loud sample is not possible.

Of the 
588 
NVSS 
sources, 
485, i.e.  about 
22\% 
of the entire AGN sample, have $R > 10$ and 
hence are radio-loud. This fraction is a lower limit because taking into account
the detection limit of 2\,mJy  in the NVSS and the optical brightness of the faint AGN the 
sample of radio non-detections could contain further radio-loud sources. The frequency
distribution of $R$  peaks around $R \approx 100$. 
All but four of the combined NVSS/GB6 matches are 
radio-loud. For these sources the mean radio spectral index is $\alpha_{\rm r} = +0.1$. 
Note that assuming a steep radio spectral index for a truly flat-spectrum source 
leads to an underestimated radio flux. 
Such a source could then fall into the group of
radio-quiet AGN, hence leading to a decrease of the radio-loud fraction.  

For the optically selected quasars from the LBQS Hooper et al. (\cite{Hooperetal96}) found 
a radio-loud fraction of $\sim10$\%.  Hooper et al. furthermore give fractions of 11\% and 
of 19\% for quasars from the EMSS and from the Palomar-Green survey (PG) using radio 
data from Stocke et al. (\cite{Stockeetal91}) and Kellermann et al. 
(\cite{Kellermannetal89}), respectively, extrapolated to 8.4\,GHz. 
Our result of $\approx$22\% radio-loud AGN is similar to the PG sample, but 
twice as high as the LBQS and, in particular, the EMSS radio-loud fraction.  
The difference from the LBQS 
might be due to selection effects, as this optically
selected sample contains mostly quasars and covers a much larger
redshift range than the X-ray selected samples, which contain a much
larger fraction of Seyfert galaxies. 
The reason for the difference from the EMSS is not as 
obvious since RASS and EMSS AGN have a comparable redshift distribution. However, 
one has to take into account that the EMSS is a serendipitous survey containing X-ray
sources found in the vicinity of bright X-ray sources. This might introduce a bias in the 
sample composition which could contribute to the differences in the radio-loud AGN 
fraction relative to an unbiased all-sky survey.

\begin{figure} 
\resizebox{\hsize}{!}{\includegraphics[angle=-90,bbllx=50pt,bblly=30pt,bburx=555pt,bbury=720pt,clip=]{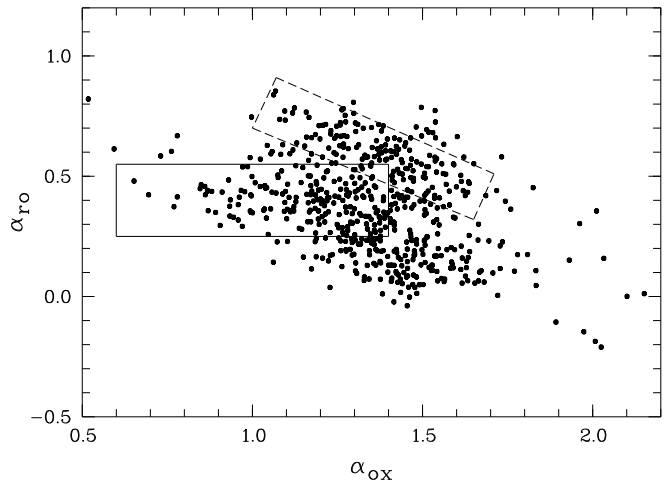}}
\caption[]{Continuum slopes $\alpha_{\rm ox}$ and $\alpha_{\rm ro}$ for AGN. The index 
$\alpha_{\rm ro}$ refers to a radio frequency of 4.85\,GHz. The solid and dashed boxes showing
the regions occupied by the two types of BL Lac objects, HBL and LBL, respectively,  were
adopted from Brinkmann et al. (\cite{B97}).
}
\label{alphaox}
\end{figure}

\begin{figure} 
\resizebox{\hsize}{!}{\includegraphics[angle=0,bbllx=20pt,bblly=40pt,bburx=550pt,bbury=550pt,clip=]{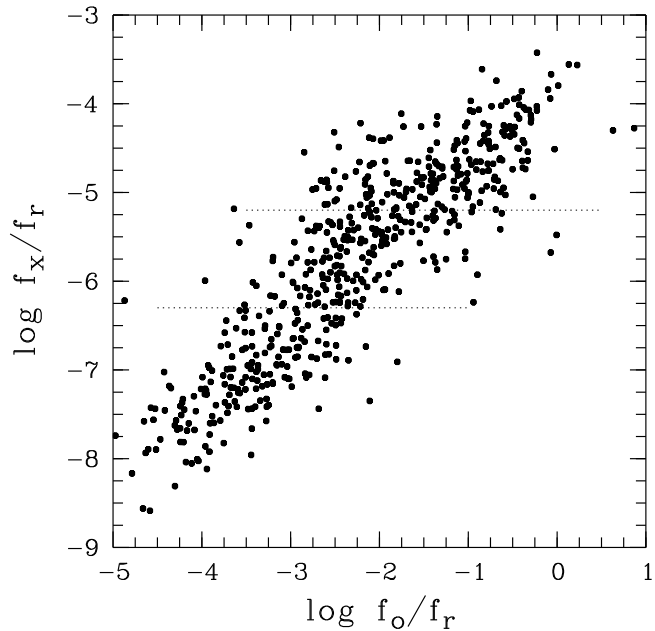}}
\caption[]{Flux ratios for  AGN. The dashed lines represent limits above and below of which HBL
and LBL/HPQ, respectively, are found (limits adopted from Brinkmann et al. \cite{B97}).
}
\label{fluxratio}
\end{figure}

Using  4.85\,GHz radio fluxes, $f_{\rm r}$, either from GB6 or extrapolated from NVSS,
monochromatic optical fluxes 
at 2500\AA , $f_{\rm o}$, and monochromatic X-ray fluxes 
at 2\,keV, $f_{\rm x}$, for photon index $\Gamma = 2.5$ (see below) we calculated the 
continuum slopes 
$\alpha_{\rm ox} = -\log (f_{\rm x}/f_{\rm o})/2.605$ and 
$\alpha_{\rm ro} = \log (f_{\rm r}/f_{\rm o})/5.38 $
as defined by Tananbaum et al. (\cite{Tananbaumetal79})
and flux ratios $f_{\rm o}/f_{\rm r}$ and  $f_{\rm x}/f_{\rm r}$. Fig. \ref{alphaox}  shows 
the $\alpha_{\rm ox}-\alpha_{\rm ro}$ diagram 
for the radio-detected AGN from the HRC.  

A special class of radio-loud AGN are the BL Lac objects. Recent surveys aiming 
specifically at  this class of objects were carried out e.g. by Laurent-Muehleisen et al. 
(\cite{Laurent98}, \cite{Laurent99}),  Rector et al. (\cite{Rector00}), and 
Beckmann et al. (\cite{Beckmannetal02}). Padovani \& Giommi (\cite{PadovaniGiommi95}) 
suggested that BL Lacs can be separated into the subgroups of low-energy peaked (LBL) and 
high-energy peaked (HBL) objects. In the $\alpha_{\rm ox}-\alpha_{\rm ro}$ 
diagram we have indicated the regions in which these two types of BL Lacs are found. 
The parameter limits were taken from  Brinkmann et al. (\cite{B97}). 

The flux ratios are depicted in  Fig. 
\ref{fluxratio} in a $\log(f_{\rm x}/f_{\rm r})$ vs. $\log(f_{\rm o}/f_{\rm r})$ diagram. 
As before the regions 
containing  LBLs and HBLs 
are indicated. 
The limits for the flux ratio of $\log(f_{\rm x}/f_{\rm r})$ are plotted as dashed lines 
in Fig. \ref{fluxratio} and were adopted from Fig. 11 in 
Brinkmann et al. (\cite{B97}). As discussed by Brinkmann et al. the class separation is more
evident in the flux ratio diagram. The number of HBL candidates with 
$\log(f_{\rm x}/f_{\rm r}) > -5.2$ is 
222. 
Below $\log(f_{\rm x}/f_{\rm r}) = -6.3$ 
186 
LBL candidates and candidates for highly polarized quasars (HPQ) are found. 
Which of the candidates are BL Lacs cannot be decided with our low-resolution
prism spectra. The spectroscopic confirmation requires  higher spectral re\-so\-lution. This
has partly been done in spectroscopic follow-up studies of BL Lac candidates 
from the HRC by Bade et al. (\cite{Badeetal98b}) and Beckmann et al.
(\cite{Beckmannetal02}) who studied a sample of nearly 100 BL Lac objects.
Although the total fraction of BL Lacs in the HRC
therefore remains uncertain at this time we can 
at least conclude that obviously less than 10\% of the AGN in our sample are HBL-type 
BL Lacs. 

\subsubsection{X-ray photon index $\Gamma$}
\label{gamma}
The X-ray hardness ratios $HR1$ and $HR2$ can be used to obtain some information about the
shape of the energy distributions in the 0.1-2.4\,keV range for AGN 
(and coronal emitters, see Sect. \ref{tcor}). 
For AGN we approximated the continuum shape $f(E)$ by a power law $f(E) \propto E^{-\Gamma}$
with the photon energy $E$ and the photon index $\Gamma$. Galactic foreground absorption due
to neutral hydrogen was taken into account according to Morrison \& McCammon
(\cite{MorrisonMcCammon83}) with \nh\ fixed at the value obtained from Dickey \& Lockman  
(\cite{DickeyLockman90}) (see Sect. \ref{fxfopt}).  

We computed hardness ratios as defined in Sec. \ref{defhard} 
from the model spectra by adding up the computed count rates in the 
corresponding energy channels and minimized the function
\begin{equation}
\Phi^2 = \sum_{k=1}^{2}\frac{(HRk_{\rm obs} - HRk_{\rm mod})^2}{\sigma (HRk)^2} 
\label{phi}
\end{equation}
with the observed and model hardness ratios $HRk_{\rm obs}$  and $HRk_{\rm
mod}$, respectively, and $k =1, 2$ (Schartel et al. \cite{Scharteletal96}). The model 
spectra were computed with EXSAS.
For each object $i$ the error of $\Gamma_i$ was estimated from $\Phi^2 = \Phi_{\rm min}^2
+1$ (one parameter problem, e.g. Lampton et al. \cite{Lamptonetal76}) as $\sigma_i
= {\rm max}(|\pm\Delta \Gamma|)$. 

\begin{figure} 
\resizebox{\hsize}{!}{\includegraphics[angle=-90,bbllx=40pt,bblly=40pt,bburx=540pt,bbury=780pt,clip=]{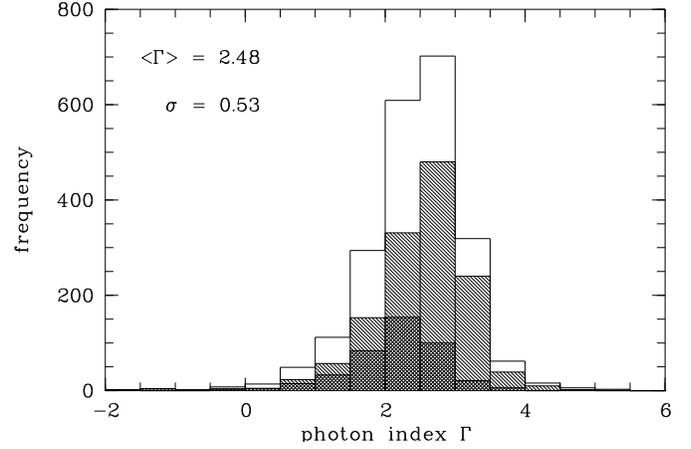}}
\caption[]{Photon index $\Gamma$ of the 2215 AGN.   
The open histogram depicts the full sample. The mean value is $\Gamma = 2.48\pm0.03$
and the intrinsic width of the distribution is $\sigma = 0.53\pm0.03$. 
The left hashed and the double-hashed histograms represent the radio-quiet and 
radio-loud AGN, respectively.
}
\label{photon}
\end{figure}

The histogram with the frequency distribution of the 
photon indices is displayed in Fig. \ref{photon}.
The mean photon index $\Gamma_m$ of the distribution was derived by means of a
maximum-likelihood estimator. We assumed a Gaussian shape for the distribution with
an intrinsic width $\sigma_{\Gamma}$. 
For $n$ data points the likelihood function is given by

\begin{equation}
 {\cal L} = \prod_{i=1}^{n} \frac{1}{\sqrt{2\,\pi (\sigma_{\Gamma}^2+
 \sigma_i^2})}e^{-\frac{1}{2}\frac{(\Gamma_m - \Gamma_i)^2}{(\sigma_{\Gamma}^2+ \sigma_i^2)}} 
\label{likeli}
\end{equation}
(Maccacaro et al. \cite{Maccacaroetal88}) with the individual photon indizes $\Gamma_i$ and their errors $\sigma_i$ for each object $i$
from Eq. \ref{phi}.
The test statistic $S = -2\,\ln {\cal L}$ was minimized with respect to $\Gamma_m$ and $\sigma_{\Gamma}$. 
Errors for these parameters were obtained from $S = S_{\rm min}+2.3$ (two parameter problem, e.g. 
Lampton et al. \cite{Lamptonetal76}). Results are summarized in Table \ref{gammatab}.
For the complete sample of 2215 AGN we thus obtained $\Gamma_m = 2.48\pm0.03$ and a intrinsic 
width $\sigma_{\Gamma} = 0.53\pm0.03$. 
For the radio-loud AGN we found  
$\Gamma_m = 2.21\pm0.05$ and  $\sigma_{\Gamma} = 0.48\pm0.05$. 
For the radio-quiet sample
the result was $\Gamma_m = 2.56\pm0.03$ and  $\sigma_{\Gamma} = 0.50\pm0.05$.
The results show that the X-ray continuum slope of radio-loud sources is significantly 
steeper than that of radio-quiet sources. The intrinsic widths of the distributions, 
however, are equal.

These findings are in very good agreement with the results of Schartel et al.
(\cite{Scharteletal96}). They studied the power-law photon indizes of a sample of 102 bright
quasars. For the soft energy band observed with ROSAT (0.1\,-2.4\,keV) they derived mean 
photon indizes of $\langle\Gamma\rangle_{rl} = 2.23\pm0.07$ and $\langle\Gamma\rangle_{rq} =
2.54\pm0.04$ for radio-loud and radio-quiet quasars, respectively. 
The slopes  from ROSAT 
were found to be significantly steeper than those determined from observations
in the harder energy range, e.g. with EINSTEIN (0.3-3.5\,keV), EXOSAT (2-10\,keV) and Ginga
(2-20\,keV). The 
slopes from ROSAT and EINSTEIN differ by
0.66 and 0.68 for radio-loud and radio-quiet quasars, respectively. 
These differences were explained by Schartel et al. as being due to the shape of the 
energy distribution. They suggested that a curved shape of the spectrum caused by an
excess in the soft energy band can explain the steeper slope observed in 
soft range with ROSAT. 

\begin{table}
\caption[]{Photon index for AGN power-law X-ray spectra.  
}   
\tabcolsep5mm
\begin{tabular}{lll}
\noalign{\smallskip}
\hline
\noalign{\smallskip}
 & $\Gamma_m$ & $ \sigma_{\Gamma}$ \\
\noalign{\smallskip}
\hline
\noalign{\smallskip}
all AGN & $2.48\pm0.03$&$0.53\pm0.03$\\
radio-loud & $2.21\pm0.05$&$0.48\pm0.05$\\
$R > 10$&&\\
radio-quiet & $2.56\pm0.03$&$0.50\pm0.05$\\
$R_{\rm lim} \le 10$&&\\
\noalign{\smallskip}
\hline
\noalign{\smallskip}
\end{tabular}
\label{gammatab}
\end{table}

\subsection{{\bf Continuum properties of stellar sources}}
\label{tcor}
\begin{figure} 
\resizebox{\hsize}{!}{\includegraphics[angle=-90,bbllx=40pt,bblly=40pt,bburx=560pt,bbury=780pt,clip=]{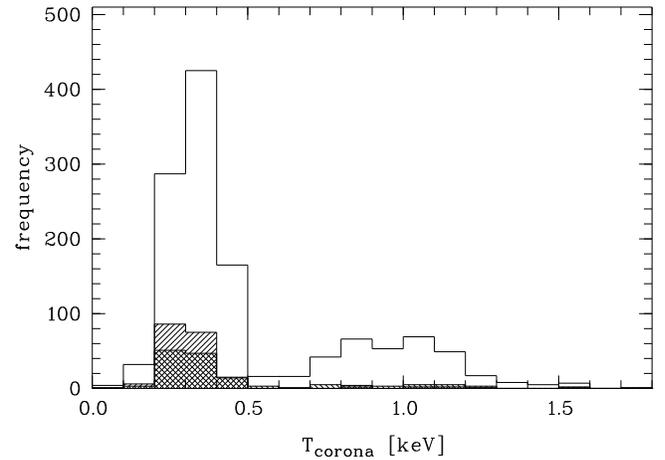}}
\caption[]{Coronal temperatures of stellar sources of spectral types M (right hashed),
K (left hashed), and of F-G and  bright stars (open histogram). 
}
\label{tcorona}
\end{figure}

For coronal emitters of classes F-G/bright, K and M  star we calculated coronal temperatures 
using the same method as before for the calculation of
the photon index for AGN. For the spectrum we assumed a one-temperature Raymond-Smith model. 
Model parameter was the coronal temperature, $T_{\rm cor}$. Galactic \nh\ was kept fixed at
1\,10$^{18}$\,\col . Furthermore, solar abundances were  assumed. The resulting temperatures 
are plotted as histograms in Fig. \ref{tcorona}. 

The frequency distribution for F-G/bright stars is bimodal with a minimum around 
0.6\,keV. It shows a dominant peak at 0.34\,keV, i.e. 4.0\,10$^6$\,K with 73\% of the
stars in the two classes below 0.6\,keV. A second peak 
appears at 1\,keV, i.e. 11.6\,10$^6$\,K with the remaining 27\% of stars above 0.6\,keV. 
K and M stars 
have temperatures around 0.30\,keV, i.e.  
3.5\,10$^6$\,K. There are also a few K and M stars in the high temperature range. The fraction
of high temperature coronae is, however, smaller for later spectral types. For K and M stars 
16\% and 7\%, respectively, have temperatures above 0.6\,keV.

The positional cross-correlation with the NVSS was also carried out for stellar counterparts. 
In 18 cases we found a radio source within the 90\% radio error circle of 10\arcsec\ 
around the optical position. 
The majority of these stars are bright stars, 2 are M stars and 1 is a K star.

\subsection{X-ray $\log N - \log S$ distributions}
\begin{figure} 
\resizebox{\hsize}{!}{\includegraphics[angle=-90,bbllx=40pt,bblly=40pt,bburx=560pt,bbury=775pt,clip=]{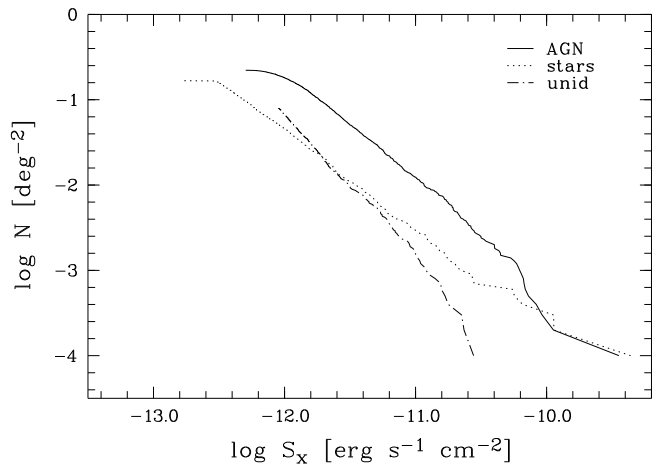}}
\caption[]{$\log N - \log S$ distributions for AGN, stars, and unidentified sources. 
}
\label{logns}
\end{figure}

In Fig. \ref{logns} $\log N - \log S$ distributions  are shown for AGN, stars, 
and unidentified sources. 
In order to derive the slopes of the distributions we assumed a cumulative
distribution function $N(>S) = k\,S^{-\alpha}$ and used 
the maximum-likelihood method described by Crawford et al. (\cite{Crawfordetal70}) to 
estimate $\alpha$. 

For the total AGN sample  
we obtained $\alpha = 1.40\pm0.05$ for the flux range $2.5\,10^{-12} \le S \le 
4.0\,10^{-11}$\,\flux\ for which the distribution is linear. 
The slope for the radio-quiet AGN is $\alpha = 1.41\pm0.07$ and for the radio-loud AGN  
$\alpha = 1.27\pm0.09$. Within the uncertainties the slopes are thus compatible with each
other and with  the Euclidean slope of 3/2.
Below  $2.5\,10^{-12}$\,\flux\ the slope becomes flatter possibly indicating 
incompleteness. 
For the LSW sample Krautter et al. (\cite{Krautteretal99}) derived a slope of
$1.41\pm0.09$ which agrees well with our result. 
Both, the LSW and our HRC slope, appear
flatter than the slope derived from the EMSS for which
Della Ceca et al. (\cite{DellaCecaetal92})  obtained $\alpha = 1.61\pm0.06$  for 
$S< 10^{-11}$\,\flux . However, the differences are on a level of $\la2\sigma$ only.
Note that inclusion of all unidentified sources would only insignificantly steepen 
our slope of the total sample to $\alpha = 1.43\pm0.05$. 
The $\log N - \log S$ distribution for unidentified sources is also shown in 
Fig. \ref{logns}.
The slope is $\alpha = 1.62\pm0.12$, hence also compatible with the Euclidean slope of 3/2.

For stellar sources we obtained  $\alpha = 1.19\pm0.03$ for the flux interval 
$4\,10^{-13} \le S \le 4\,10^{-11}$\,\flux .  The slope is thus
significantly flatter than Euclidean. The subgroup of coronal emitters (F-G/bright, K, 
and M stars) yields a slightly yet insignificantly steeper slope of $\alpha = 1.22\pm0.04$.
Inclusion of all unidentified sources in the stellar group would steepen the slope to  
$\alpha = 1.54\pm0.05$, i.e. compatible with the Euclidean slope.

\section{Summary}
\label{sum}
We presented a catalogue of  optical identifications of 5341 northern high-galactic 
latitude X-ray sources from the RASS-BSC. The catalogue contains $\sim51$\%
extragalactic sources  and $\sim 31$\% galactic stellar X-ray emitters. The majority of
extragalactic sources are  active galactic nuclei (Sy, QSOs, BL
Lacs) ($\sim$41\% of all sources). Stellar sources are mostly coronal emitters.  
In 3\% of the sources no optical counterpart was visible on our
blue direct plates. 
Another 15\% of the sources remain unidentified.

In general, the source statistics of our catalogue is in good agreement with previous studies
of smaller samples from the RASS. 
Some small differences in the frequency distribution
of the individual object classes seem to be present, though.

A surprisingly large fraction of about 20\% of AGN in the 
HRC is radio-loud. This is about twice as much as in the LBQS and in the EMSS.
Our analysis of the  X-ray spectra of  AGN showed that radio-loud AGN have 
significantly harder X-ray continua than radio-quiet AGN.

The HRC is a valuable data base for the selection of complete samples of soft X-ray emitters
for detailed follow-up studies.

\acknowledgements{
This work was supported by DARA under grant Verbundforschung  
BMBF\,\,50\,\,OR\,\,9606\,\,0.  We would like to thank the referee Dr. Perlman for his
constructive comments which helped to improve the paper.
The Hamburg Quasar Survey (HQS) was supported by the
Deutsche Forschungsgemeinschaft through grants Re\,353/11-1,2,3  and Re\,353/22-1,2,3.
The ROSAT project was supported by the Bundesministerium 
f\"ur Bildung und Wissenschaft and the Max-Planck-Gesellschaft. 
The Digitized Sky Survey was produced at the Space Telescope Science 
Institute under U.S. Government grant NAG W-2166. 
}

\appendix
\section{Sample spectra}
\label{spectra}
Sample spectra for the various spectral classes were extracted from the digital data 
base of the HQS prism plates. For the spectral classes of stars (F-M stars, CVs, 
white dwarfs) and AGN (including QSOs and BL Lacs) we selected objects with known 
characteristics from various catalogues. For each object we give the classification
according to our object classes and in parentheses the catalogue classification and 
brightness. 
 
\begin{figure*} 
\centering
\includegraphics[width=17cm,angle=0,bbllx=65pt,bblly=445pt,bburx=555pt,bbury=780pt,clip=]{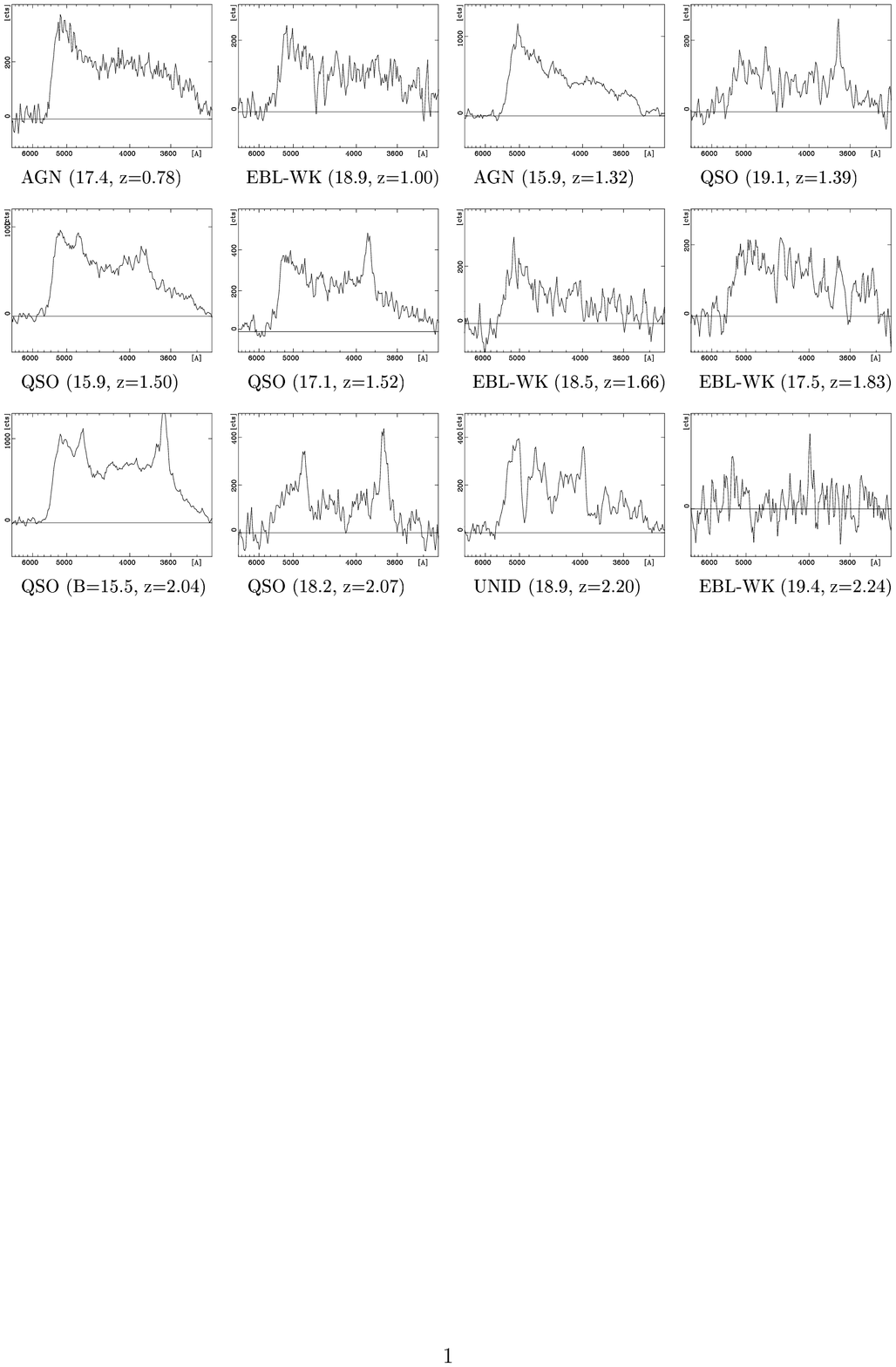}
\caption[]{Sample spectra of QSOs with various redshifts and brightnesses.
}
\label{qsospec}
\end{figure*}

\begin{figure*} 
\centering
\includegraphics[width=17cm,angle=0,bbllx=65pt,bblly=670pt,bburx=555pt,bbury=780pt,clip=]{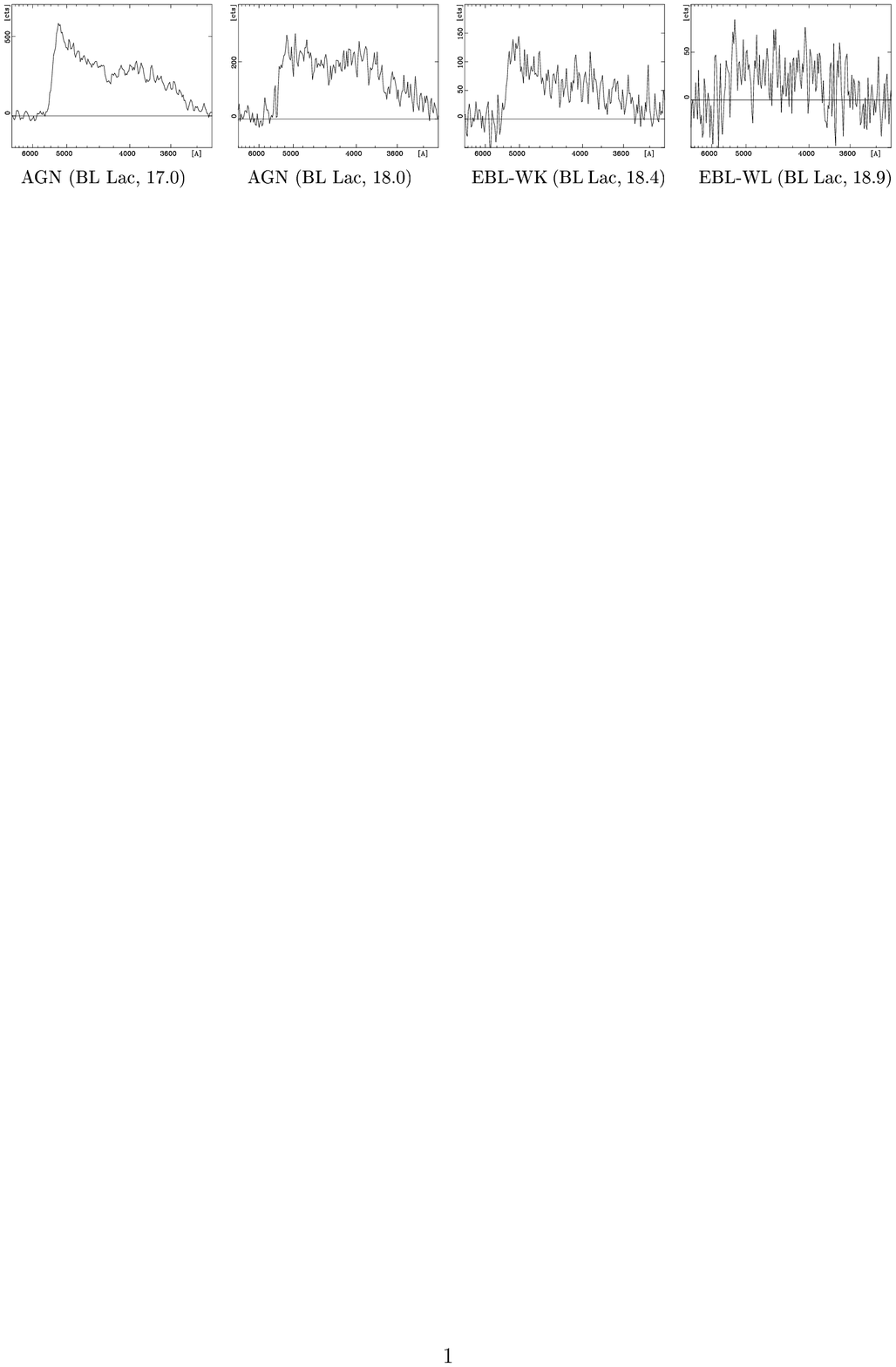}
\caption[]{Sample spectra of BL Lac objects with various  brightnesses.
}
\label{bllacspec}
\end{figure*}

\begin{figure*} 
\centering
\includegraphics[width=17cm,angle=0,bbllx=65pt,bblly=670pt,bburx=555pt,bbury=780pt,clip=]{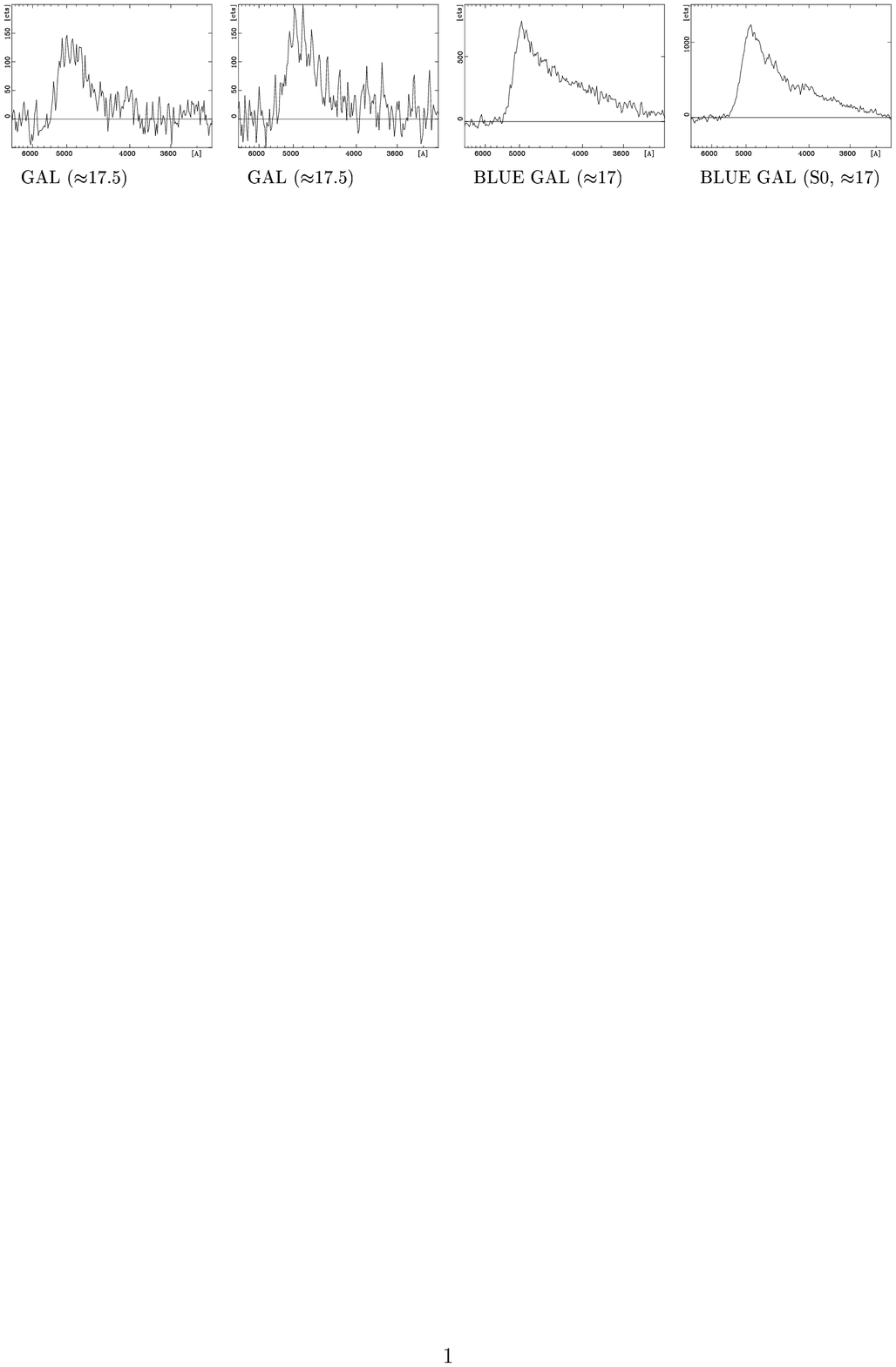}
\caption[]{Sample spectra of galaxies.
}
\label{galspec}
\end{figure*}

\begin{figure*} 
\centering
\includegraphics[width=17cm,angle=0,bbllx=65pt,bblly=305pt,bburx=555pt,bbury=745pt,clip=]{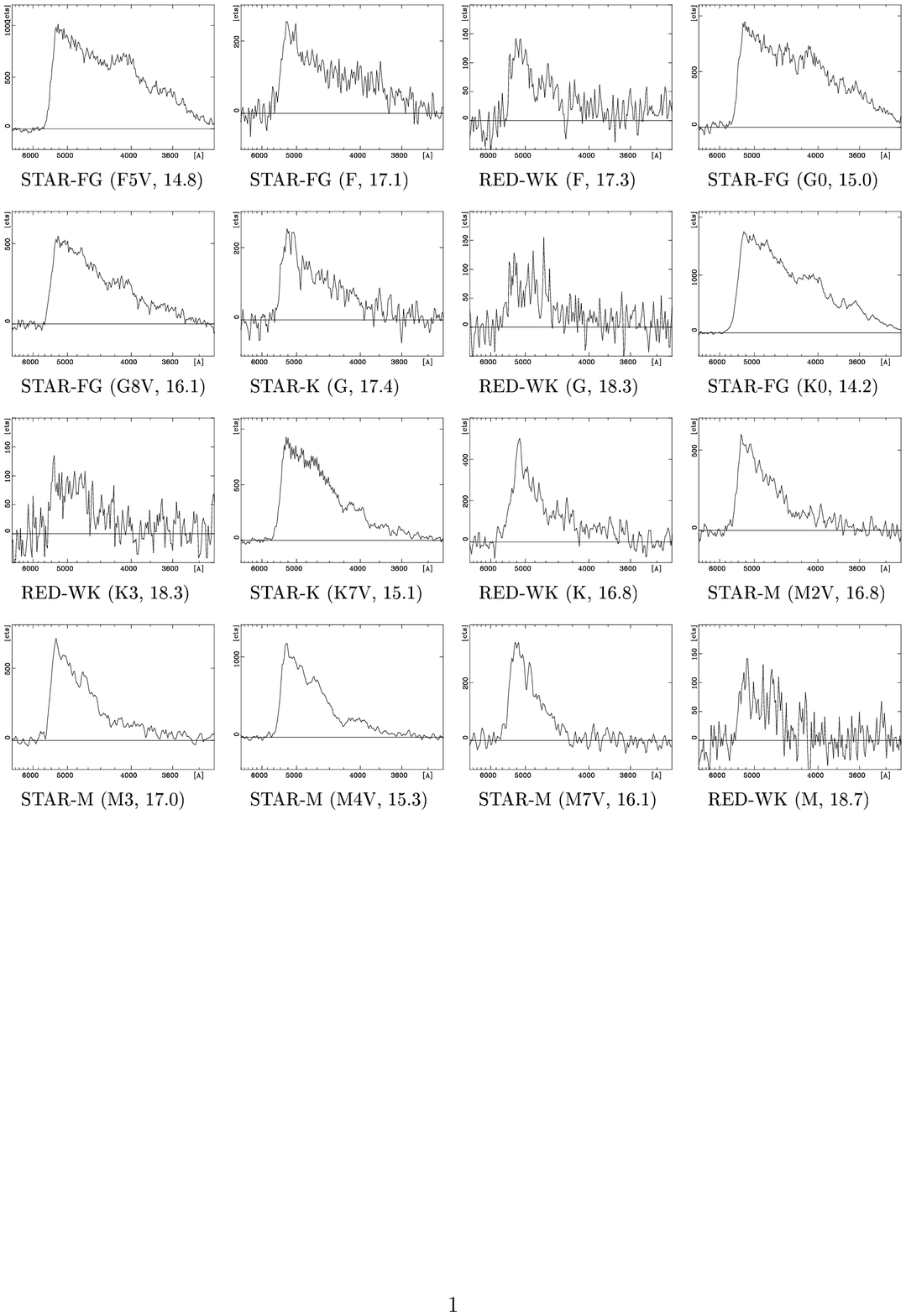}
\caption[]{Sample spectra of F, G, K and M type stars.
}
\label{fmspec}
\end{figure*}

\begin{figure*} 
\centering
\includegraphics[width=17cm,angle=0,bbllx=65pt,bblly=670pt,bburx=555pt,bbury=780pt,clip=]{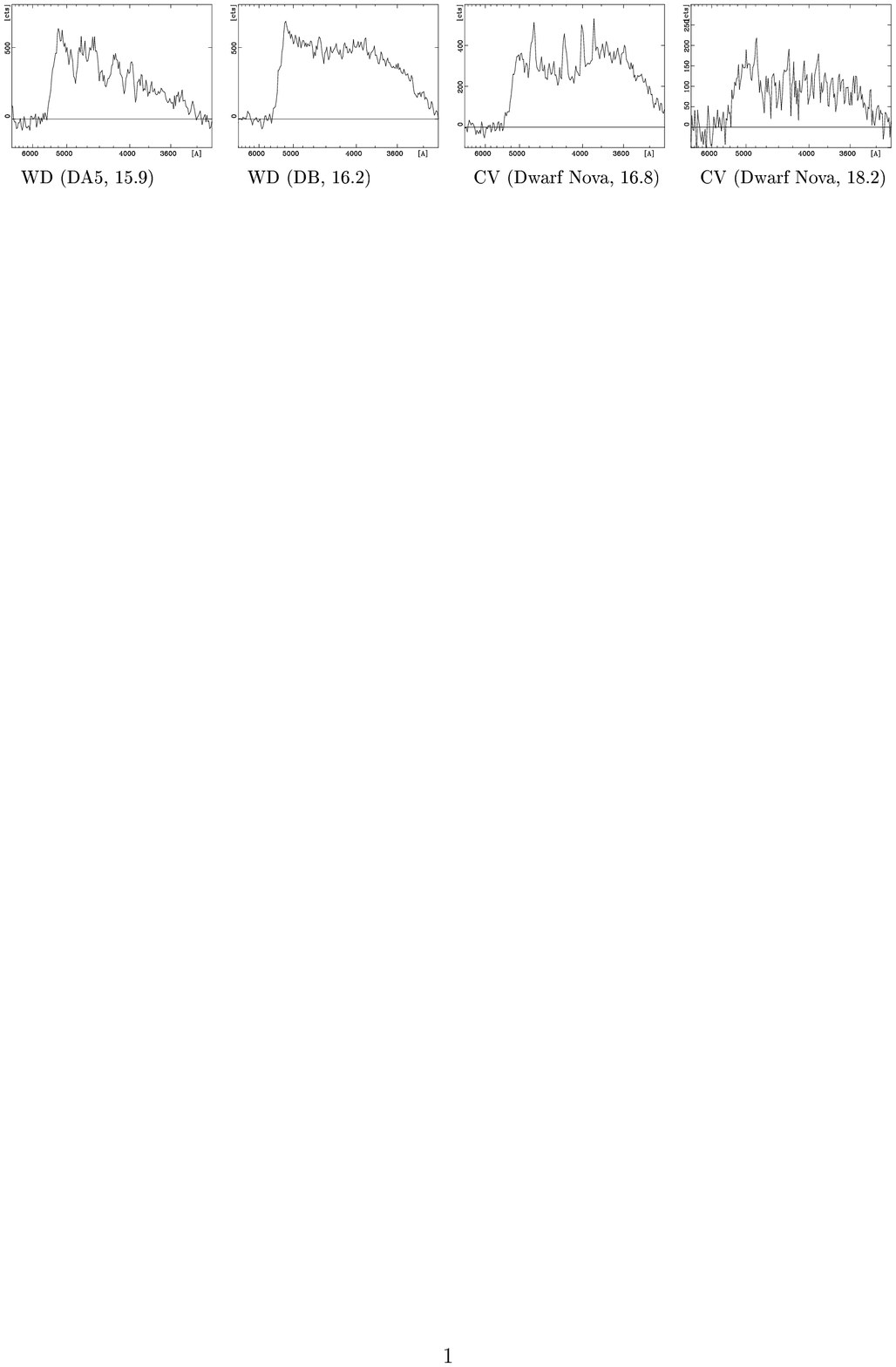}
\caption[]{Sample spectra of white dwarfs and cataclysmic variables.
}
\label{wdspec}
\end{figure*}

\end{document}